\documentclass[reprint,superscriptaddress,amsmath,amssymb,twocolumn,amsfonts,floatfix, longbibliography, 10pt]{revtex4-2}
\usepackage{graphicx}
\usepackage{epstopdf}
\usepackage[T1]{fontenc}
\usepackage[utf8]{inputenc}
\usepackage{amsbsy}
\usepackage{gensymb}

\usepackage[dvipsnames]{xcolor}
\definecolor{deepfuchsia}{rgb}{0.76, 0.33, 0.76}
\definecolor{electricpurple}{rgb}{0.75, 0.0, 1.0}
\usepackage[colorlinks=true,linktoc=page,linkcolor=blue,citecolor=magenta,urlcolor=electricpurple]{hyperref}
\usepackage{orcidlink}

\usepackage{multibib}  

\newcites{Supp}{References for Supplementary Information}

\usepackage{titlesec} 

\titleformat{\section}
  {\raggedright\bfseries\large} 
  {\thesection}{1em}{}

\titleformat{\subsection}
  {\raggedright\bfseries\large} 
  {\thesubsection}{1em}{}

\titlespacing*{\section}
  {0pt} 
  {5pt} 
  {0pt} 

\titlespacing*{\subsection}
  {0pt} 
  {3pt} 
  {0pt} 

\usepackage[normalem]{ulem}
\usepackage{textcomp} 

\newcommand{\beq}{\begin{equation}}
\newcommand{\eeq}{\end{equation}}
\newcommand{\bea}{\begin{eqnarray}}
\newcommand{\eea}{\end{eqnarray}}

\newcommand{\fig}[1]{Fig.~\ref{#1}}

\usepackage{mathtools}

\DeclarePairedDelimiter\ket{\lvert}{\rangle}
\DeclarePairedDelimiterX\braket[2]{\langle}{\rangle}{#1 \delimsize\vert #2}

\renewcommand\thesection{\arabic{section}}
\renewcommand{\thesubsection}{\alph{subsection}}

\begin{document}

\title{\textrm{Topological superconductivity in hourglass Dirac chain metals (Ti, Hf)IrGe}}

\author{{Pavan Kumar Meena}\,\orcidlink{0000-0002-4513-3072}}\thanks{These authors contributed equally to this work}
\affiliation{Department of Physics, Indian Institute of Science Education and Research Bhopal, Bhopal, 462066, India}
\author{{Dibyendu Samanta}\,\orcidlink{0009-0004-3022-7633}}
\thanks{These authors contributed equally to this work}
\affiliation{Department of Physics, Indian Institute of Technology, Kanpur 208016, India}
\author{Sonika Jangid}
\affiliation{Department of Physics, Indian Institute of Science Education and Research Bhopal, Bhopal, 462066, India}
\author{Roshan Kumar Kushwaha}
\affiliation{Department of Physics, Indian Institute of Science Education and Research Bhopal, Bhopal, 462066, India}
\author{Rhea Stewart}
\affiliation{ISIS Facility, STFC Rutherford Appleton Laboratory, Didcot OX11 0QX, United Kingdom}
\author{Adrian D. Hillier}
\affiliation{ISIS Facility, STFC Rutherford Appleton Laboratory, Didcot OX11 0QX, United Kingdom}
\author{{Sudeep Kumar Ghosh}\,\orcidlink{0000-0002-3646-0629}}
\email[]{skghosh@iitk.ac.in}
\affiliation{Department of Physics, Indian Institute of Technology, Kanpur 208016, India}
\author{{Ravi Prakash Singh}\,\orcidlink{0000-0003-2548-231X}}
\email[]{rpsingh@iiserb.ac.in}
\affiliation{Department of Physics, Indian Institute of Science Education and Research Bhopal, Bhopal, 462066, India}

\begin{abstract}
Realizing topological superconductivity in stoichiometric materials is a key challenge in condensed matter physics. Here, we report the discovery of ternary germanide superconductors, $M$IrGe ($M$ = Ti, Hf), as prime candidates for topological superconductivity, predicted to exhibit nonsymmorphic symmetry-protected hourglass Dirac chains. Using comprehensive thermodynamic and muon-spin rotation/relaxation ($\mu$SR) measurements, we establish these materials as conventional bulk type-II superconductors with transition temperatures of 2.24(5) K for TiIrGe and 5.64(4) K for HfIrGe, featuring a full gap and preserved time-reversal symmetry. First-principles calculations reveal striking topological features in $M$IrGe, including hourglass-shaped bulk dispersions and a Dirac chain- a ring of fourfold-degenerate Dirac points protected by nonsymmorphic symmetry. Each Dirac point corresponds to the neck of the hourglass dispersion, while the Dirac chain gives rise to drumhead-like surface states near the Fermi level. Additionally, nontrivial $\mathbb{Z}_2$ topology leads to isolated Dirac surface states with helical spin textures that disperse across the Fermi level, forming an ideal platform for proximity-induced topological superconductivity. The coexistence of conventional bulk superconductivity, symmetry-protected hourglass topology, and helical spin-textured surface states establishes $M$IrGe as a rare and robust platform to realize topological superconductivity, opening new avenues for next-generation quantum technologies.
\end{abstract}
\keywords{}
\maketitle

\section{Introduction}
Symmetry-protected topological phases have revolutionized quantum materials research, leading to the discovery of Dirac and Weyl semimetals in materials like Na$_3$Bi, Cd$_3$As$_2$, TaAs, LaAlGe, and MoTe$_2$~\cite{Armitage2018,ong2021experimental,Lv2021}. These systems exhibit unique transport signatures and robust surface states driven by non-trivial topology of their bulk electronic structures~\cite{ong2021experimental,wang2017quantum}. A key breakthrough has been the recognition of non-symmorphic crystal symmetries, which enforce essential band crossings and entangle multiple bands, giving rise to unconventional fermionic excitations such as hourglass fermions~\cite{wang2016hourglass, alexandradinata2016topological}. Characterized by hourglass-shaped bulk dispersions, these fermions are stabilized by nonsymmorphic symmetries like glide mirrors and screw axes, ensuring robust surface states across different crystal orientations~\cite{wang2016hourglass, alexandradinata2016topological}. Unlike topological insulators, which rely solely on time-reversal symmetry~\cite{hasan2010colloquium}, hourglass metals benefit from additional symmetry protection, making them promising candidates for advanced electronics due to their spin-momentum locking and high spin-Hall conductivity~\cite{wang2016hourglass,Singh2018,gao2020r,Li2018,ma2017experimental}. 

In parallel, topological superconductors have emerged as a focal point in quantum research, particularly for their potential in topological quantum computing~\cite{qi2011topological, nayak2008non, sato2017topological}. While efforts to realize topological superconductivity have often relied on proximity effects or doping in materials like Bi$_2$Se$_3$~\cite{wang2012coexistence, wang2013fully, kriener2011bulk, sasaki2011topological} and SnTe~\cite{sasaki2012odd}, these approaches face challenges such as interface effects and lattice mismatches. This has spurred the search for intrinsic topological superconductors in stoichiometric materials~\cite{yan2013large, guan2016superconducting} that combine high superconducting transition temperatures ($T_c$) with isolated topological surface states at the Fermi level. Although candidates like Au$_2$Pb~\cite{xing2016superconductivity} and PdTe$_2$~\cite{noh2017experimental} have been identified, their low $T_c$ values underscore the rarity of ideal materials. Bulk superconducting topological metals, which intertwine topology, correlations, and novel superconducting ground states, offer a promising platform for realizing intrinsic topological superconductivity and exploring exotic quantum phenomena.

\begin{figure*}[!htb]
\includegraphics[width=\textwidth]{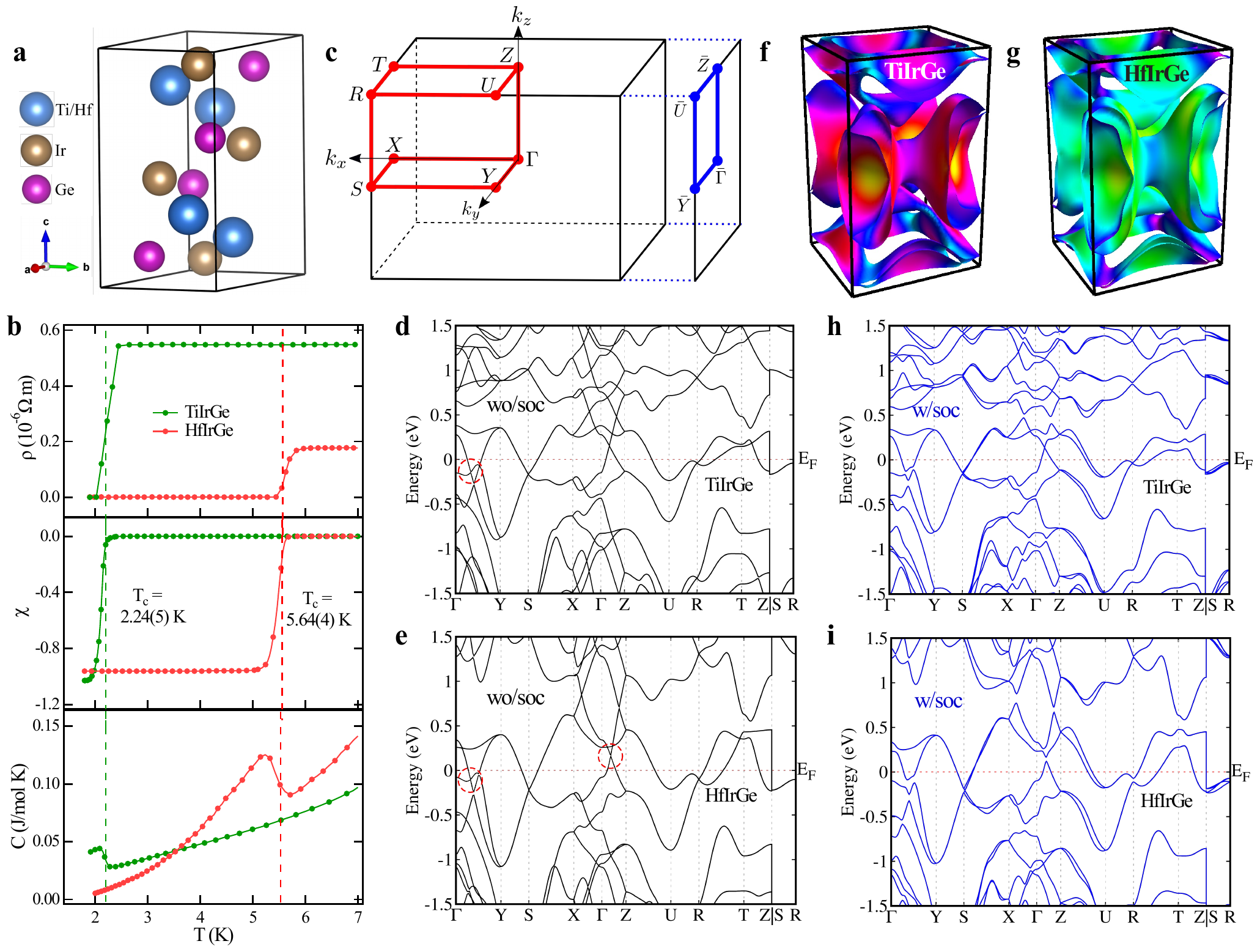}
\caption {Bulk superconductivity and Electronic band structure of $M$IrGe ($M$ = Ti, Hf): a) Schematic of the crystal structure of $M$IrGe ($M$ = Ti, Hf). b) Temperature variation of resistivity (top), magnetization (middle) in zero-field-cooled warming and field-cooled cooling modes at an external magnetic field $\mu_0 H = 1.0$ mT, and specific heat (bottom), showing superconductivity of $M$IrGe ($M$ = Ti, Hf). c) Bulk Brillouin zone (BZ) and its projection onto the (100) surface, with red dots and blue lines indicating high-symmetry points and paths, respectively. d,e) Electronic band structures of TiIrGe and HfIrGe without spin-orbit coupling (SOC). f,g) Combined Fermi surfaces with SOC for TiIrGe and HfIrGe, respectively. The band crossings that lead to glide mirror symmetry-protected nodal rings are shown by red circles. h,i) Electronic band structures with SOC for TiIrGe and HfIrGe, respectively.}
\label{fig:structure}
\end{figure*} 
Equiatomic ternary silicide and germanide materials~\cite{gupta2015review, morozkin1999crystallographic, landrum1998tinisi, welter1993crystallographic, subba1985structure} are known for their diverse crystal structures and remarkable properties, including complex magnetism and Kondo lattice behavior. Recent breakthroughs have revealed unconventional superconductivity with time-reversal symmetry breaking (TRSB) in Weyl nodal-line semimetals such as La(Pt, Ni)Si and LaPtGe~\cite{shang2022spin,kp2018superconducting, kp2020probing}, while nodeless superconductivity has been observed in ThIrSi~\cite{tay2023nodeless} through $\mu$SR studies. TRSB has also been reported in centrosymmetric TiNiSi-type structures such as (Nb, Ta)OsSi~\cite{ghosh2022time}, but not in (Zr, Hf)IrSi~\cite{panda2019probing, bhattacharyya2019investigation} compounds. Furthermore, ScRuSi and ZrRhSi have been extensively studied both theoretically and experimentally to demonstrate fully gapped conventional superconductivity~\cite{uzunok2020first, panda2024probing}. However, the bulk superconductivity of $M$IrGe (M = Ti, Hf)~\cite{wang1987crystal} compounds, which have nonsymmorphic crystal symmetries and are isostructural to U(Rh, Co)Ge~\cite{hattori2012superconductivity} known for the coexistence of ferromagnetism and superconductivity with possible triplet pairing, remains largely unexplored.

In this paper, we investigate the ternary germanides $M$IrGe ($M$ = Ti, Hf)~\cite{wang1987crystal} as promising candidates for topological superconductors using a combination of experimental and theoretical techniques. Substituting Ti (3$d$) with Hf (5$d$) enables exploration of the effects of enhanced spin-orbit coupling. Resistivity, magnetization, and specific heat measurements confirm bulk conventional type-II superconductivity, while $\mu$SR studies reveal an isotropic s-wave superconducting gap with preserved TRS. First-principles calculations and symmetry analysis reveal hourglass-type bulk dispersions, where the necks form Dirac rings protected by nonsymmorphic symmetries, leading to drumhead-like surface states. These materials also host well-separated Dirac topological surface states with helical spin textures due to nontrivial $\mathbb{Z}_2$ topology. These robust, symmetry-protected topological surface states disperse across the Fermi level and remain distinct from the bulk states, providing an ideal platform to investigate the interplay between nonsymmorphic symmetry-protected topological features and superconductivity.

\section{Results}

\noindent \textbf{Structural characterization and bulk superconductivity:}
Polycrystalline samples of the ternary germanides $M$IrGe ($M$ = Ti, Hf) were synthesized by the arc melting method. Powder X-ray diffraction (XRD) measurements confirm that $M$IrGe ($M$ = Ti, Hf) crystallizes in an orthorhombic TiNiSi-type structure [Figure~\ref{fig:structure}a], with nonsymmorphic space group symmetry Pnma (No. 62) and point group $D_{2h}$ (see Supplemental material (SM)). The temperature dependence of resistivity, magnetic susceptibility, and specific heat of $M$IrGe ($M$ = Ti, Hf), as shown in Figure~\ref{fig:structure}b, confirms bulk superconductivity, exhibiting a sharp resistivity drop, full diamagnetic screening in zero-field cooled warming at an applied field of 1.0 mT, and a pronounced jump in specific heat at the transition temperature ($T_c$). The observed $T_c$ values are approximately 2.24(5) K for TiIrGe and 5.64(4) K for HfIrGe. The normal-state resistivity, fitted to the parallel resistor model, decreases with temperature, showing metallic behavior with RRR values of 5.14(6) for TiIrGe and 6.52(7) for HfIrGe. The FCC and ZFCW curves indicate flux pinning and type-II superconductivity. Field-dependent magnetization and temperature-dependent resistivity/magnetization measurements enabled the estimation of the lower and upper critical fields by fitting the temperature dependence of these critical fields with the Ginzburg-Landau model (GL) (see SM for details). The lower critical field values $H_{c1}(0)$ are 5.6 (1) mT for TiIrGe and 36.4 (1) mT for HfIrGe, while the upper critical fields $H_{c2}(0)$ are 0.68(1) and 1.36(1) T (from magnetization) and 0.71(1) and 2.04(1) T (from resistivity) for TiIrGe and HfIrGe, respectively. Using the $H_{c1}$(0) and $H_{c2}$(0) values, the obtained coherence lengths $\xi_{GL}$ for TiIrGe and HfIrGe are 22.0(2) nm and 15.5(6) nm, respectively, while the penetration depths $\lambda_{GL}$ are 279.1(5) nm and 92.6(3) nm, yielding GL parameters $\kappa_{GL}$ of 12.6(8) and 5.9(5), with thermodynamic critical fields H$_{c}$ estimated at 38.7(3) mT and 166.6(2) mT, indicating strong type-II superconductivity. The low value of the Maki parameters (0.18(1) for TiIrGe and 0.15(1) for HfIrGe) suggests that orbital limiting effects dominate over the Pauli limiting effect in their superconductivity (see SM for detailed calculation). Normal state-specific heat data are fitted to the Debye-Sommerfeld relation, the resulting Debye temperatures $\theta_{D}$ are 399(5) and 301(2) K, and the density of states at the Fermi level $D_{C}(E_{F})$ is 4.21(6) and 3.15(6) states $eV^{-1}$ $f.u.^{-1}$, respectively. The electron-phonon coupling constants $\lambda_{e-ph}$, calculated from McMillan's equation, are 0.48(9) and 0.65(7) for TiIrGe and HfIrGe, suggesting weak-coupling superconductivity. Electronic specific heat $C_{el}(T)$ is well fitted with the fully gapped weak-coupling BCS model (see SM), yielding superconducting gaps of $\Delta/k_{B}T_{c}$ = 1.47(2) and 2.04(2) for TiIrGe and HfIrGe, respectively.

\begin{figure*}
\includegraphics[width=\textwidth]{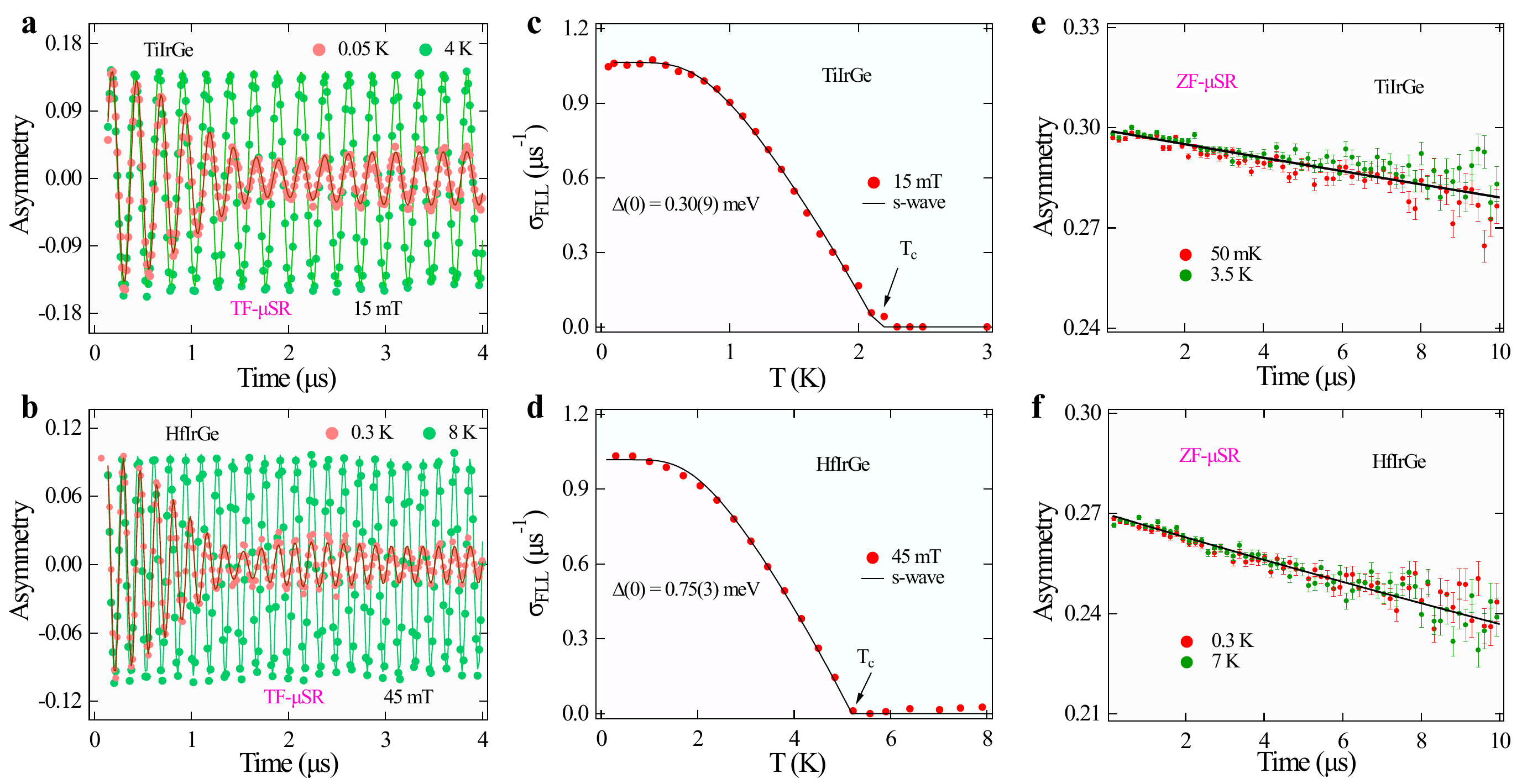}
\caption{Microscopic muon spin rotation and relaxation ($\mu$SR) results of $M$IrGe ($M$ = Ti, Hf): a,b) show TF-$\mu$SR asymmetry spectra at 15 and 45 mT for Ti and Hf-based compounds in the superconducting and normal states, with solid lines representing the fit using eq. \ref{Gaussian oscillatory}. c,d) Temperature-dependent relaxation rate fitted with an s-wave model in the superconducting states for both samples. e,f) display ZF-$\mu$SR asymmetry spectra below and above $T_c$, with solid lines representing the respective fit for TiIrGe and HfIrGe.}
\label{fig:muSR}
\end{figure*}

\noindent \textbf{Electronic band structure:} To elucidate the characteristics of the normal state of $M$IrGe ($M$ = Ti, Hf), we performed detailed calculations of the electronic band structure calculations from first principles using density functional theory (DFT) within the generalized gradient approximation (GGA)~\cite{Perdew1996}. The 3D bulk and (100) 2D surface Brillouin zones of $M$IrGe ($M$ = Ti, Hf) are schematically shown in Figure~\ref{fig:structure}c. The calculated electronic band structures without the effects of spin-orbit coupling (SOC) are shown in Figure~\ref{fig:structure}d,e for TiIrGe and HfIrGe, respectively. We note that multiple dispersive bands intersect the Fermi level, forming various electron and hole pockets, indicating the multi-band nature of $M$IrGe ($M$ = Ti and Hf). These two isostructural compounds show similar band dispersions, featuring band crossing points along the $\Gamma-Y$ and $\Gamma-Z$ paths in the Brillouin zone, indicated by red circles. The low-energy bands near the Fermi level in TiIrGe are predominantly composed of Ti-3d, Ir-5d, Ge-4p, and Ir-5p orbitals, whereas in HfIrGe, they consist of Hf-5d, Ir-5d, Ge-4p, Ir-5p, and Hf-5p orbitals, both in descending order of significance. These compositions align with the calculated orbital-resolved projected density of states (DOS) (see SM for details). SOC induces significant splittings in the band structures near the Fermi level, as shown in Figure~\ref{fig:structure}d,e for TiIrGe and HfIrGe, respectively, with SOC. The estimated maximum band splittings caused by SOC are along the $RT$-direction and are given by $\sim 100$ meV for TiIrGe and $\sim 150$ meV for HfIrGe, reflecting the substantial contributions to the bands from the Ti $3d$ and Hf $5d$ orbitals, respectively. The SOC band structures of both compounds reveal distinctive hourglass-type dispersions along the S-X and S-R high-symmetry directions, for example, as illustrated in detail later. The combined Fermi surface with parallel sheets of TiIrGe and HfIrGe with SOC is shown in Figure~\ref{fig:structure}f,g, respectively. We note that there are several parallel Fermi surface sheets spread across the Brillouin zone, raising the possibility of dominant interband superconducting pairing~\cite{ghosh2022time,Weng2016}.

\begin{figure*}[!htb]
\includegraphics[width=\textwidth]{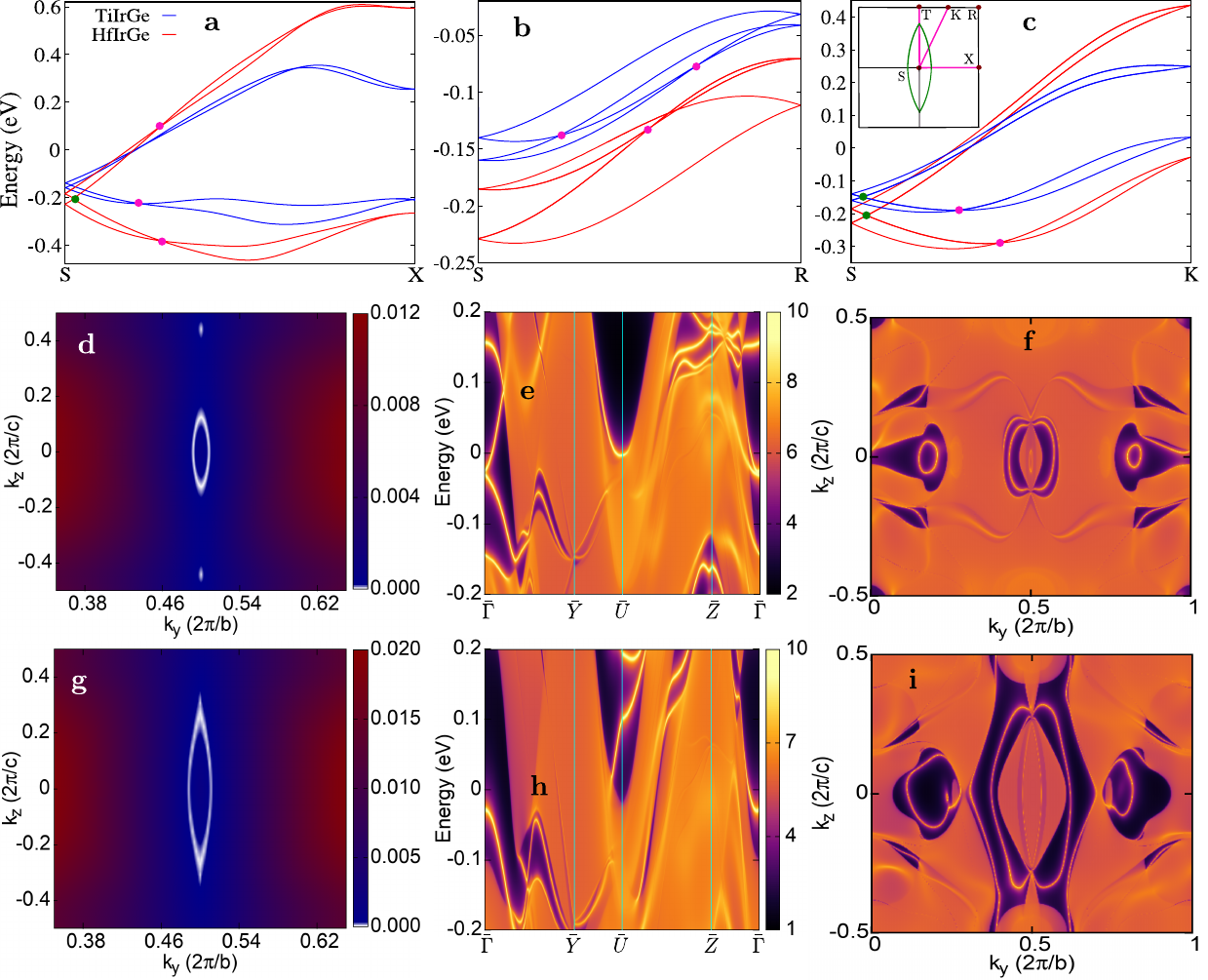}
\caption {\label{fig:hourglass} Hourglass dispersions, Dirac rings and surface states of $M$IrGe ($M$ = Ti, Hf) with SOC: a-c) Hourglass-type band dispersions for TiIrGe and HfIrGe along the high-symmetry directions $S-X$, $S-R$ and $S-K$ (TiIrGe in blue, HfIrGe in red). Here, $K$ represents the midpoint between $T$ and $R$ as shown in the inset of (c). Type-I and type-II Dirac fermions are marked by green and pink dots, respectively. The inset depicts the schematic of the fourfold degenerate Dirac ring formed by neck points (green dots) of the hybrid hourglass dispersion on the $k_x = \pi$ plane. d,g) Distribution of the hourglass Dirac ring (white) surrounding point S for TiIrGe and HfIrGe, respectively. The color scale indicates the local gap between crossing bands. e,h) Surface state spectrum along the high-symmetry paths in the projected (100) 2D surface Brillouin zone for TiIrGe and HfIrGe, respectively. f,i) Surface Fermi arcs (constant energy slice of the spectrum) for TiIrGe at energy -0.138 eV and for HfIrGe at energy -0.130 eV, respectively.}
\end{figure*} 

\noindent \textbf{Muon spin rotation and relaxation ($\mu$SR) results:} Comprehensive $\mu$SR measurements were conducted using the MuSR spectrometer at the ISIS Pulsed Neutron and Muon Source in the UK \cite{hillier2022muon}. The $\mu$SR technique, renowned for its exceptional sensitivity arising from the muons significant magnetic moment and large gyromagnetic ratio, provides invaluable insights into the ground-state properties of superconductors. Results from the transverse-field (TF) and longitudinal-field (LF) configurations are discussed below.

\noindent \textbf{TF-$\mu$SR:} TF-$\mu$SR measurements, with an external magnetic field applied perpendicular to the muon spin polarization, probe the superconducting gap structure, revealing insights into carrier density in the superconducting state through the formation of a flux line lattice (FLL) in fields between the lower ($H_{c1}$) and upper ($H_{c2}$) critical fields. Asymmetry spectra were recorded at temperatures from 50 mK to 4 K and 0.3 K to 8 K for the TiIrGe and HfIrGe samples under different applied transverse magnetic fields. Figure~\ref{fig:muSR}a,b show representative spectra for the TiIrGe and HfIrGe samples above and below $T_{c}$ in fields of 15 and 45 mT, respectively. Above $T_{c}$, there is a small normal state relaxation due to randomly oriented nuclear moments, whereas below $T_{c}$, the formation of the FLL in the mixed state results in an inhomogeneous magnetic field that causes a dramatic increase in the damping. The time evolution of the TF asymmetry can be well described by the Gaussian-damped oscillatory function given as \cite{weber1993magnetic, maisuradze2009comparison};
\begin{equation}
A(t) =  A_{1} e^{-\sigma^{2} t^{2}/2} \cos(\gamma_{\mu}B_{1}t+\phi) + A_{2}\cos(\gamma_{\mu}B_{2}t+\phi),
\label{Gaussian oscillatory}
\end{equation}
where $A_{1}$ and $A_{2}$ denote the initial asymmetries of the sample and the non-relaxing background from the silver sample holder, respectively, while the corresponding local magnetic fields sensed by the muons are $B_{1}$ and $B_{2}$.$\gamma_{\mu}/2 \pi$ = 135.5 MHz/T is the muon gyromagnetic ratio, and $\phi$ is the phase of the initial muon spin polarization with respect to the detector. Above $T_c$, the internal magnetic field equals the applied field, while in the superconducting state, the Meissner effect reduces the internal field (see SM). The superconducting relaxation rate ($\sigma_{FLL}$) is calculated using $\sigma_{FLL} = \sqrt{\sigma^{2} - \sigma^{2}_{N}}$, where $\sigma$ is the total relaxation rate, while $\sigma_{N}$ accounts for the temperature-independent normal state relaxation due to the randomly oriented nuclear spins. The $\sigma_{N}$ values are 0.049(2) and 0.033(6) $\mu s^{-1}$ for TiIrGe and HfIrGe compounds, respectively. The temperature-dependent relaxation $\sigma_{FLL}(T)$, directly proportional to the penetration depth and superconducting fluid density, is fitted with the equation;
\begin{equation}
\frac{\sigma_{FLL}(T)}{\sigma_{FLL}(0)} = \frac{\lambda^{-2}(T)}{\lambda^{-2}(0)}= 1 + 2 \int_{\Delta(T)}^{\infty} \left(\frac{\partial f}{\partial E} \right) \frac{E dE}{\sqrt{E^{2}-\Delta(T)^{2}}},
\label{s-wave}
\end{equation}
where $\lambda(0)$ is the London penetration depth, $f = [1+ e^{(E/k_{B}T)}]^{-1}$ is the Fermi distribution function and $\Delta (T)$ is the BCS superconducting gap function, defined by $\Delta(T) = \Delta_{0}\tanh[1.82(1.018((T_{c}/T)-1))^{0.51}]$ \cite{prozorov2006magnetic}. The $\sigma_{FLL}(T)$ were well described by the isotropic BCS s-wave model [Figure~\ref{fig:muSR}c,d], providing superconducting gaps of $\Delta(0)$ = 0.30(9) and 0.75(2) meV, with normalized gaps $\Delta(0)/k_{B}T_{C}$ = 1.66(7) and 1.68(2) for TiIrGe and HfIrGe, respectively~\cite{bhattacharyya2019investigation, panda2019probing, panda2024probing, ghosh2022time}. Brandt has reported that, for a superconductor in the case of a low magnetic field such as $H << H_{c2}$, the field-dependent relaxation rate $\sigma_{sc}$ (see SM), associated with penetration depth $\lambda(T)$, can be fitted with the equation given below;
\begin{equation}
\sigma_{sc} (\mu s^{-1}) = 4.854 \times 10^{4} (1-h)[1+1.21(1-\sqrt{h})^3] \lambda^{-2}
\label{penetrationdepth}
\end{equation}
with $h = H/H_{c2}(0)$ the reduced field. The obtained London penetration depth $\lambda^{\mu SR}(0)$ values are 2731(6) and 2461(9) \text{\AA} for TiIrGe and HfIrGe, respectively. These values are comparable to the other structurally similar compounds (Zr, Hf)IrSi \cite{panda2019probing, bhattacharyya2019investigation, panda2024probing, ghosh2022time}. Notably, the $\lambda^{\mu SR}(0)$ value for TiIrGe aligns closely with magnetization estimates, while HfIrGe shows a significant discrepancy. This divergence, observed in other compounds as well, may reflect a genuine difference in determining penetration depth from the lower critical field (in the Meissner state) and superfluid density (in the vortex state)~\cite{anand2018superconductivity, mandal2021superconducting, kataria2023superconducting}. The SM provides the electronic properties and includes the Uemura plot.

\noindent \textbf{ZF-$\mu$SR:} ZF-$\mu$SR is used to detect the precession of muon spins in local magnetic fields in the sample, which can reveal the presence or absence of time-reversal symmetry in the superconducting state. ZF-$\mu$SR asymmetry spectra for polycrystalline $M$IrGe ($M$ = Ti and Hf) are similar above and below T$_{c}$ as shown in Figure~\ref{fig:muSR}e,f, ruling out TRSB due to the presence of spontaneous magnetic field within the detection limits (up to 1 $\mu$T). A similar trend with a lower asymmetry value is also reported in silicides \cite{panda2019probing, tay2023nodeless}. The absence of coherent oscillations or fast decays in the spectra indicates that there is no magnetic order or fluctuations, despite the possibility of TRSB-induced magnetization increasing relaxation rates~\cite{ghosh2020recent}. In non-magnetic materials, the depolarization of muon spins in zero field is mainly determined by the randomly oriented nuclear magnetic moments. The ZF depolarization data was fitted with a Lorentzian function;
\begin{equation}
G_{ZF}(t) = A_{0}(t)\exp{(-\lambda t)} +A_{bg},
\label{Lorentzian function}
\end{equation}
where $A_{0}$ is the initial asymmetry corresponding to the sample, and $A_{bg}$ considers the background asymmetry associated with muon stopping in the silver sample holder, which is nearly temperature-independent. The relaxation rate $\lambda$, associated with nuclear moments, shows negligible temperature dependence in the ZF-$\mu$SR asymmetry data, suggesting that there is no evidence of TRSB in the $M$IrGe samples. However, the breaking of TRS in other similar compounds raises questions about the role of spin-orbit coupling in these materials.

\noindent \textbf{Topology of electronic band structure:} Both $M$IrGe ($M$ = Ti, Hf) have rich topological features in their electronic band structures that are protected by nonsymmorphic glide mirror symmetries. The complete symmetry of the crystal structure of $M$IrGe ($M$ = Ti and Hf) can be generated by the three key symmetry operations: inversion symmetry $\mathcal{P}:(x,y,z)\rightarrow(-x,-y,z)$, mirror symmetry $\mathcal{M}_y:(x,y,z)\rightarrow(x,-y+\frac{1}{2},z)$, and glide mirror symmetry $\mathsf{G}_x:(x,y,z)\rightarrow(-x+\frac{1}{2},y+\frac{1}{2},z+\frac{1}{2})$. The two isostructural compounds show similar band dispersions without SOC, featuring band crossing points along the $\Gamma-Y$ and $\Gamma-Z$ paths in the Brillouin zone, as indicated by the red circles in Figure~\ref{fig:structure}d,e. These crossings give rise to a nodal ring encircling the $\Gamma$ point in the $k_x = 0$ plane (see SM for details), protected by the glide mirror symmetry $\mathsf{G}_x$. When SOC is included [Figure~\ref{fig:structure}h,i], the nodal ring surrounding the $\Gamma$ point in the $k_x = 0$ plane fully gaps out, leading to band inversion near the $\Gamma$ point in both compounds and consequently nontrivial topology. Although the $M$IrGe compounds ($M$ = Ti, Hf) do not have a global band gap across the entire Brillouin zone (BZ) with SOC, we can still define the topological invariant $\mathbb{Z}_2$ in the $k_y = 0$ plane. Our calculations reveal that $M$IrGe has a topological index of $\mathbb{Z}_2 = 1$, indicating the presence of nontrivial topological surface states on the (010) surface that have a helical spin texture, are well separated from the bulk, and disperse across the Fermi level (see SM for details).

The band structures of both compounds with SOC reveal distinctive hourglass-type dispersions along the $S-X$, $S-R$ and $S-K$ (where $K$ denotes the midpoint between $T$ and $R$) high-symmetry paths, as shown in Figure~\ref{fig:structure}h,i. Further analysis reveals several distinct topological features: glide mirror $\mathsf{G}_x$ protected hourglass dispersions, a Dirac ring around the S-point in the $k_x = \pi$ plane, and nontrivial topological surface states. The zoomed versions of the unique hourglass-shaped dispersions near the Fermi levels protected by the glide mirror symmetry $\mathsf{G}_x$ along the $S-X$, $S-R$, and $S-K$ high-symmetry paths are shown in {Figure}~\ref{fig:hourglass}a-c. 

First we examine the hourglass dispersion along the $S-R$ high symmetry path [i.e. along ($\pi,\pi,k_z$), where $-\pi < k_z < \pi$]. This path in the $k_x=\pi$ plane remains invariant under the glide mirror operation $\mathsf{G}_x$. Along $S-R$, $\mathsf{G}^2_x=\mathcal{T}_{011}\Bar{E}=e^{-ik_z}$, where $\Bar{E}$ denotes a $2\pi$ spin rotation and $\mathcal{T}_{011}$ translates $(x,y,z)$ to $(x,y+b,z+c)$. Consequently, the eigenvalue $g_x$ of $\mathsf{G}_x$ must be $\pm e^{-ik_z/2}$. For a Bloch state $\ket{\phi_n}$, its Kramer's partner $\mathcal{P}\mathcal{T}\ket{\phi_n}$ satisfies $\mathsf{G}_x(\mathcal{PT}\ket{\phi_n})=g_x(\mathcal{PT}\ket{\phi_n})$, sharing the same eigenvalue $g_x$. At the time-reversal invariant momenta (TRIM) points $S$ and $R$, $\mathsf{G}_x(\mathcal{T}\ket{\phi_n})=g_x(\mathcal{T}\ket{\phi_n})$. Given the centrosymmetric crystal structures of $M$IrGe ($M$ = Ti, Hf), $\mathsf{G}_x(\mathcal{P}\ket{\phi_n})=g_x(\mathcal{P}\ket{\phi_n})$. At $S$ $(\pi, \pi, 0)$, the states $\ket{\phi_n}$, $\mathcal{T}\ket{\phi_n}$, $\mathcal{P}\ket{\phi_n}$, and $\mathcal{PT}\ket{\phi_n}$ then form a degenerate quartet with $g_x=\pm 1$ ($g_x=-1$ has a lower energy than $g_x=+1$). At $R$ $(\pi, \pi, \pi)$, $g_x=\pm i$, where the Kramer's partners have opposite eigenvalues, forming a quartet with two states with $g_x=+i$ and the other two with $g_x=-i$. This symmetry-enforced eigenvalue redistribution between the points $S$ and $R$ necessarily leads to symmetry-protected band crossing, manifesting as the characteristic hourglass dispersion shown in Figure~\ref{fig:hourglass}b. Since these band crossings are enforced by nonsymmorphic glide mirror symmetry and independence from band inversion mechanism, they are very robust. Similarly, topological analysis also applies to the $S-X$ and $S-K$ high-symmetry paths leading analogously to glide mirror symmetry-protected hourglass dispersions as shown in Figure~\ref{fig:hourglass}a,c. We note that the hourglass dispersions are spread across a very large energy window crossing the Fermi level along the $S-X$ and $S-K$ directions (Figure~\ref{fig:hourglass}a,c) in contrast to the dispersions along the $S-R$ direction in a very narrow energy window below the Fermi level (Figure~\ref{fig:hourglass}b).

Figure~\ref{fig:hourglass}a-c also feature additional type-I and type-II Dirac point crossings shown by the green and pink dots, respectively. Type-I Dirac points (neck points) arise for any $S-K$ path, where $K$ is an arbitrary point along the $T-R$ direction. A similar concept also applies to the $R-X$ high symmetry path. Consequently, the neck points (green dots) of the hourglass-type dispersion in $M$IrGe ($M$ = Ti and Hf) are guaranteed and form a continuous Dirac nodal ring around the point $S$ in the $k_x = \pi$ plane [as schematically shown in the inset of Figure~\ref{fig:hourglass}c]. This hybrid hourglass-type Dirac ring, confirmed by DFT calculations and shown for TiIrGe and HfIrGe in Figure~\ref{fig:hourglass}d,g, respectively, is fundamentally determined by the nonsymmorphic space group symmetry.

Surface electronic structure analysis of TiIrGe and HfIrGe, projected onto the (100) surface Brillouin zones (Figure~\ref{fig:hourglass}e,f), unveils multiple topologically non-trivial surface states intersecting the Fermi level. These materials exhibit bulk nodal loops that manifest as characteristic drumhead-like surface states in regions where the loops project finite areas onto the surface ~\cite{Li2018}. Due to broken inversion symmetry on the surface, the drumhead surface bands show splitting~\cite{Li2018}. The states, which are fundamentally connected to the bulk Dirac rings, emerge at specific energies: -0.138 eV in TiIrGe and -0.130 eV in HfIrGe (Figure~\ref{fig:hourglass}h,i).

\section{Discussions}
The low symmetry crystal structure of $M$IrGe ($M$ = Ti and Hf) and the strong effects of SOC on their electronic band structure provide a unique opportunity to determine the symmetry of the superconducting order parameter. Using a Ginzburg-Landau theory-based symmetry analysis~\cite{ghosh2020recent} of the superconducting order parameters in the strong SOC limit within an effective single-band picture corresponding to the point group $D_{2h}$, we find that all the symmetry allowed superconducting order parameters apart from the one in the fully symmetric $^1A_1$-channel have nodes~\cite{Hillier2012}. Nonsymmorphic symmetries can also lead to additional symmetry-enforced nodes at the Brillouin Zone boundaries or zone faces~\cite{Sumita2018}. Since $M$IrGe ($M$ = Ti and Hf) shows bulk superconductivity with a single gap and no signatures of TRS breaking, the leading superconducting instability in these materials will be in the $^1A_1$ s-wave singlet channel coming mostly from phonon-mediated on-site pairing. The inherently multiband character of $M$IrGe leading to a pair of parallel Fermi surface sheets, which are close to each other throughout the Brillouin zone as shown in \fig{fig:structure}f, g, will give rise to a strong interband pairing. Further theoretical investigations using a detailed tight-binding model of the band structure of $M$IrGe, for example, with on-site singlet pairing, are necessary to calculate the novel transport properties~\cite{ma2017experimental} expected in these materials. 

We computed the phonon dispersions of $M$IrGe ($M$ = Ti, Hf) that reveal real positive phonon spectra throughout the entire Brillouin zone, confirming the dynamic stability of their crystal structures. Using phonon spectra, the superconducting transition temperature $(T_c)$ is calculated using the McMillan formula~\cite{Allen1975}, assuming phonon-mediated s-wave superconductivity in these materials. $T_c$ is determined by evaluating electron-phonon coupling matrix elements through the Eliashberg spectral function $\alpha^2 F(\omega)$~\cite{McMillan1968,Allen1972}. The cumulative electron-phonon coupling constant ($\lambda$) is given by $\lambda = 2 \int \frac{\alpha^2 F(\omega)}{\omega}  d\omega$. Using these parameters, the superconducting critical temperature $T_c$ is calculated using the modified McMillan formula~\cite{Allen1975}: $T_c = \frac{\omega_{\log}}{1.2} \exp \left[ \frac{-1.04(1+\lambda)}{\lambda - \mu^*(1+0.62 \lambda)} \right]$. $\mu^*$ is an empirical parameter that describes the screened Coulomb interaction~\cite{Morel1962} and has typical values between $0.1$ and $0.16$. The logarithmically averaged phonon frequency is $\omega_{\log}=\exp \left[ \frac{2}{\lambda} \int \frac{\alpha^2 F(\omega)}{\omega} \log (\omega) d\omega \right]$. We obtained $T_c$ for TiIrGe and HfIrGe to be $2.71$ K and $5.08$ K, respectively, assuming $\mu^*=0.10$~\cite{Morel1962}, in close agreement with the experimental results.

The symmetry-protected topological surface states in $M$IrGe ($M$ = Ti and Hf) are well separated from the bulk states and disperse across the Fermi level, enabling the emergence of a distinct surface superconducting gap through proximity coupling with the bulk superconducting condensate. This interplay establishes $M$IrGe as a promising platform to realize surface topological superconductivity. The coexistence of a distinct bulk superconducting gap and a smaller surface gap can be experimentally probed using techniques such as Andreev reflection spectroscopy, as demonstrated in YRuB$_2$~\cite{mehta2024topological}, or angle-resolved photoemission spectroscopy (ARPES), as applied to PdTe~\cite{Yang2023}.

\section{Summary and conclusion}
We systematically uncover the superconducting and topological properties of polycrystalline $M$IrGe ($M$ = Ti and Hf), demonstrating a striking interplay between conventional superconductivity and nonsymmorphic symmetry-protected hourglass topology, thereby establishing these materials as robust stoichiometric platforms for realizing topological superconductivity. X-ray diffraction confirms the TiNiSi-type orthorhombic crystal structure with nonsymmorphic space group symmetry. Bulk superconductivity is established through resistivity, magnetization, and specific heat measurements, yielding $T_c \approx$ 2.24(5) K for TiIrGe and 5.64(4) K for HfIrGe. Critical field and specific heat analysis indicate conventional type-II weak-coupling BCS superconductivity with an isotropic s-wave gap, further supported by muon spin rotation and relaxation ($\mu$SR) measurements confirming fully gapped superconductivity with preserved time-reversal symmetry. First-principles calculations reveal a rich topological landscape, featuring nonsymmorphic glide mirror symmetry-protected hybrid hourglass bulk dispersions, where the necks of these dispersions form a Dirac chain that generates drumhead-like surface states. Notably, topological surface states arising from band inversion around the $\Gamma$-point cross the Fermi level with a helical spin texture on the (100) surface, providing an ideal setting to explore topological superconductivity. The presence of these topological features in $M$IrGe can be experimentally verified, for example, using ARPES or scanning tunneling microscopy/spectroscopy (STM/STS)~\cite{Chen2022}.

As nonsymmorphic symmetry-protected hourglass Dirac chain metals that are also superconductors, $M$IrGe compounds offer exciting prospects for advanced quantum technologies. Their symmetry-protected topological surface states can enable proximity-induced superconductivity~\cite{Li2018}, potentially hosting robust Majorana modes for fault-tolerant quantum computing. The coexistence of helical spin textures and superconductivity could facilitate spin-polarized supercurrents, essential for spintronic applications~\cite{ma2017experimental,Eschrig2015}. Bulk Dirac nodal loops in $M$IrGe will lead to emergent unconventional low-energy excitations, which can give rise to novel superconducting instabilities with unconventional pairing mechanisms. Additionally, these materials can form Josephson junctions with unique current-phase relationships~\cite{Parhizgar2020}, suitable for quantum interference devices, while coupling with dissipative elements could open pathways in non-Hermitian quantum physics~\cite{Cayao2024,Shen2024}. Their Dirac nodal loops with drumhead-like surface states could also enable intriguing transport phenomena~\cite{Burkov2011, Mullen2015, Yu2017}, including anisotropic charge transport~\cite{Mullen2015}, anomalous Landau levels~\cite{Rhim2015}, and exotic optical responses~\cite{Ahn2017,Liu2018}, making them promising candidates for next-generation quantum and nanoelectronic devices.


\section{Methods}
\noindent \small{\textbf{Sample characterization:} Polycrystalline $M$IrGe ($M$ = Ti and Hf) samples were synthesized by arc melting of high-purity ($4N$) Ti or Hf, Ir, and Ge in a 1:1:1 stoichiometric ratio in an Ar atmosphere, followed by rapid cooling and multiple remelts for homogeneity. The ingots were then annealed at 850\degree C for 7 days. Phase purity and crystal structure were analyzed by powder X-ray diffraction (XRD) with Cu $K_{\alpha}$ radiation ($\lambda$ = 1.5406 $\text{\AA}$) on a PANalytical diffractometer. Low-temperature magnetization measurements were performed using a Quantum Design MPMS-3 (7T) in vibrating sample magnetometry mode. Electrical and specific heat measurements were performed with the Quantum Design PPMS 9T system using four-probe and two-tau techniques.

\noindent \textbf{$\mu$SR measurements:} The $\mu$SR measurements were conducted on the MuSR spectrometer at the ISIS pulsed neutron and muon facility, STFC Rutherford Appleton Laboratory, UK, using 64 detectors in transverse field (TF) and zero field (ZF) configurations to investigate superconducting pairing mechanisms and spontaneous internal magnetic fields. Powder samples were mounted on a silver holder with diluted GE varnish and cooled in a dilution refrigerator. The muon asymmetry signal is determined from: $A(t) = [N_{F}(t)-\alpha N_{B}(t)]/[N_{F}(t)+\alpha N_{B}(t)]$, where $N_{B}(t)$ and $N_{F}(t)$ are the number of counts in the backward and forward detector, respectively, and $\alpha$ is an experiment-specific constant determined from calibration measurements taken with a small applied transverse magnetic field. The reference provides an in-depth explanation of the technique \cite{hillier2022muon}. Data were analyzed using MANTID software.

\noindent \textbf{Electronic band structure calculations:}
We performed first-principles electronic structure calculations using density functional theory (DFT) as implemented in the QUANTUM ESPRESSO package. The Perdew-Burke-Ernzerhof (PBE) functional was employed within the generalized gradient approximation (GGA) to account for exchange-correlation effects. Projector augmented wave (PAW) pseudopotentials described the electron-ion interactions, with a plane-wave basis set expanded to a kinetic energy cut-off of 80 Ry. Brillouin zone integration utilized a $\Gamma$-centered $8 \times 10 \times 8$ Monkhorst-Pack k-point mesh for bulk calculations. Experimental lattice parameters and atomic positions, derived from the Rietveld refinement of X-ray diffraction data, were used in our calculations. We constructed a tight-binding Hamiltonian based on Maximally Localized Wannier Functions (MLWFs) using the WANNIER90 package. This Wannier-based Hamiltonian served as input for subsequent calculations of topological properties, including surface states and nodal loops, which were performed using the WANNIER TOOLS package.}


\section{Acknowledgments}
PKM and DS contributed equally to this work. PKM acknowledges the funding agency Council of Scientific and Industrial Research, Government of India, for providing the SRF fellowship award no. 09/1020(0174)/2019-EMR-I. RPS acknowledges the Science and Engineering Research Board (SERB), Government of India, for the Core Research Grant No. CRG/2023/000817 and ISIS, STFC, UK, for providing beamtime for the $\mu$SR experiments~\cite{Experiment_Runnumber}. SKG also acknowledges financial support from SERB, Government of India via the Startup Research Grant: SRG/2023/000934 and IIT Kanpur via the Initiation Grant (IITK/PHY/2022116). DS and SKG utilized the \textit{Andromeda} server at IIT Kanpur for numerical calculations. The authors acknowledge Kapildeb Dolui for discussions.



\section*{References}
\bibliography{Hourglass_Main_refs, Hourglass_suppl_refs} 

\section*{Supplementary Materials}
Supplementary Text

Figure S1 to S10

Table S1-S2

References [70-87]

\clearpage

\onecolumngrid  

\begin{center}
    {\LARGE \textbf{Supplementary Material}} \\
    \vspace{15pt}
    {\Large \textbf{Topological superconductivity in hourglass Dirac chain metals (Ti, Hf)IrGe}} \\
\end{center}

\vspace{10pt}


\setcounter{figure}{0} 

\renewcommand{\thefigure}{S\arabic{figure}}  

\setcounter{table}{0} 

\renewcommand{\thetable}{S\arabic{table}}  


In the supplementary material, we present additional details of the synthesis, physical characteristics, data analysis and band structure of the $M$IrGe ($M$ = Ti, Hf) compounds.

\section*{Band structure and topology}
We conducted ab initio electronic structure calculations using density functional theory (DFT) as implemented in QUANTUM ESPRESSO. The exchange-correlation effects were treated within the generalized gradient approximation (GGA) using the Perdew-Burke-Ernzerhof (PBE) functional. Electron-ion interactions were described by projector augmented wave (PAW) pseudopotentials. The plane-wave basis set was expanded up to a kinetic energy cutoff of 80 Ry. Brillouin zone integration employed a $\Gamma$-centered $8 \times 10 \times 8$ Monkhorst-Pack k-point mesh for bulk calculations. We used experimental lattice parameters and atomic positions derived from Rietveld refinement of X-ray diffraction data.

\begin{figure}[h]
\includegraphics[width=0.7\columnwidth,origin=b]{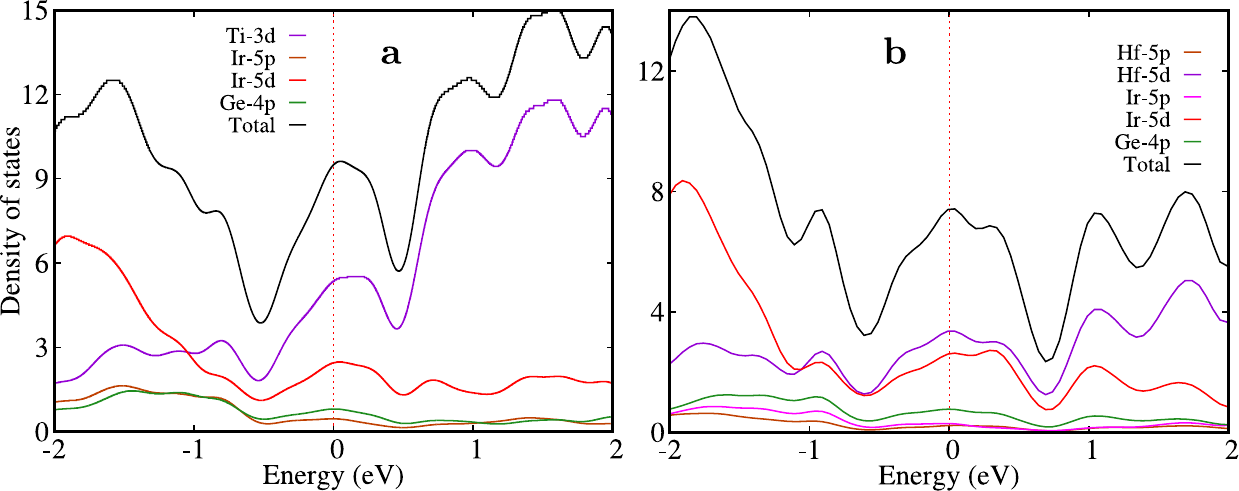}
\caption{Orbital resolved electronic density of states without spin-orbit coupling (SOC) for a) TiIrGe and b) HfIrGe.}
\label{fig:pdos}
\end{figure}

Figure \ref{fig:pdos} presents the orbital-resolved projected density of states (PDOS) calculated without spin-orbit coupling (SOC) for $M$IrGe ($M$ = Ti, Hf). In TiIrGe [Figure~\ref{fig:pdos}a], the electronic states near the Fermi level are dominated by Ti-3d orbitals, followed by significant contributions from Ir-5d and Ge-4p orbitals, with minor Ir-5p involvement. This substantial d-orbital character, particularly from Ti-3d and Ir-5d states, elucidates the observed SOC-induced band splitting in TiIrGe. For HfIrGe [Figure \ref{fig:pdos}b], the near-Fermi level states are primarily composed of Hf-5d orbitals, with notable contributions from Ir-5d, Ge-4p, Ir-5p, and Hf-5p orbitals, in descending order of significance. The pronounced presence of 5d orbitals from the heavy elements Hf and Ir accounts for the significant band splitting observed in the SOC-inclusive band structure of HfIrGe.

\begin{figure}[t]
\includegraphics[width=0.7\columnwidth,origin=b]{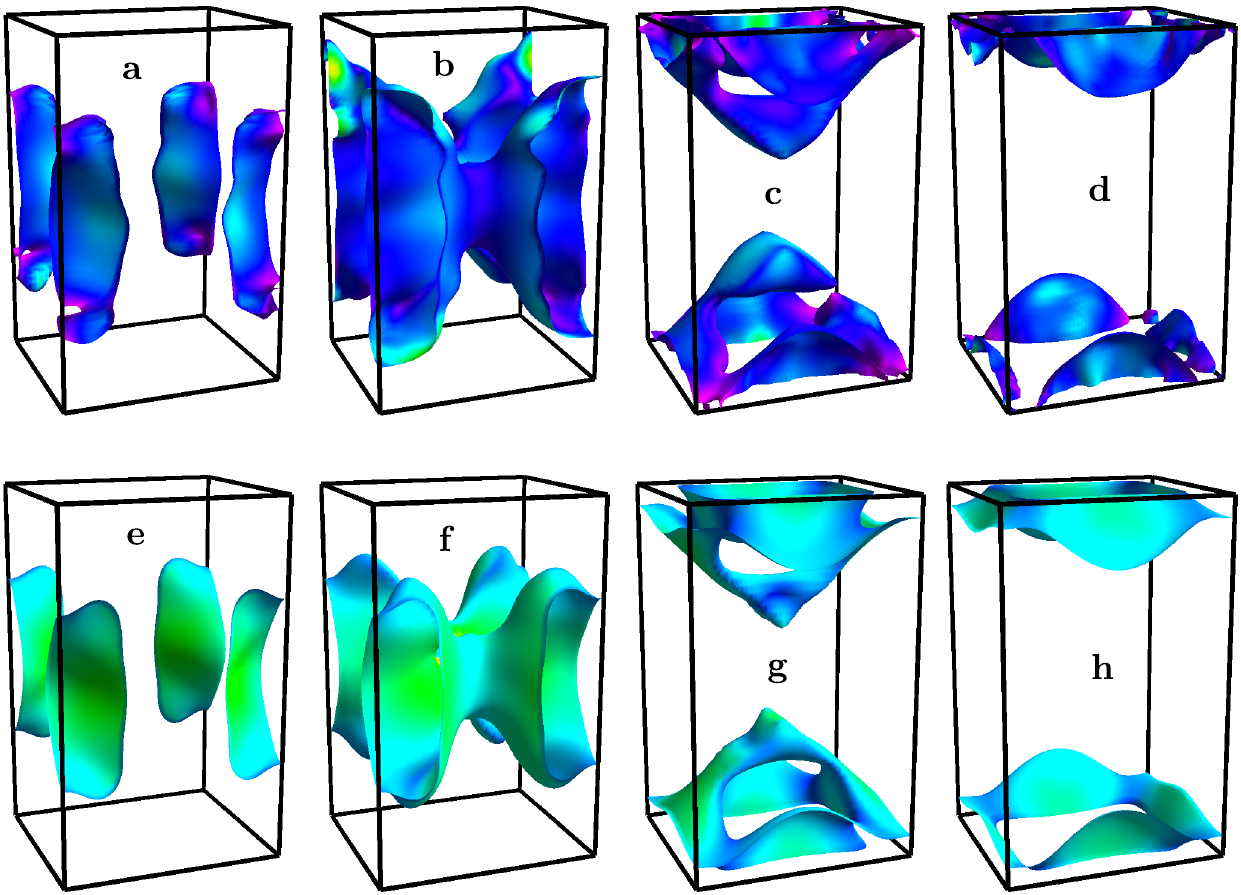}
\caption{Fermi surfaces without spin-orbit coupling: a-d) TiIrGe and e-h) HfIrGe, illustrating distinct electron and hole pockets in the first Brillouin zone.}
\label{fig:fs}
\end{figure}

To compute the Fermi surfaces of $M$IrGe ($M$ = Ti, Hf), we constructed a tight-binding Hamiltonian based on Maximally Localized Wannier Functions (MLWFs). The MLWFs were generated using the WANNIER90 software package. Figure~\ref{fig:fs} illustrates these Fermi surfaces, obtained without incorporating SOC. Both TiIrGe and HfIrGe exhibit four Fermi surface sheets each, as illustrated in Figure~\ref{fig:fs}a-d and e-h, respectively, with each sheet representing a band that crosses the Fermi level. All Fermi sheets contribute equally to the electronic density of states, underscoring the multi-band nature of both compounds.

\begin{figure}[b]
\includegraphics[width=0.6\columnwidth,origin=b]{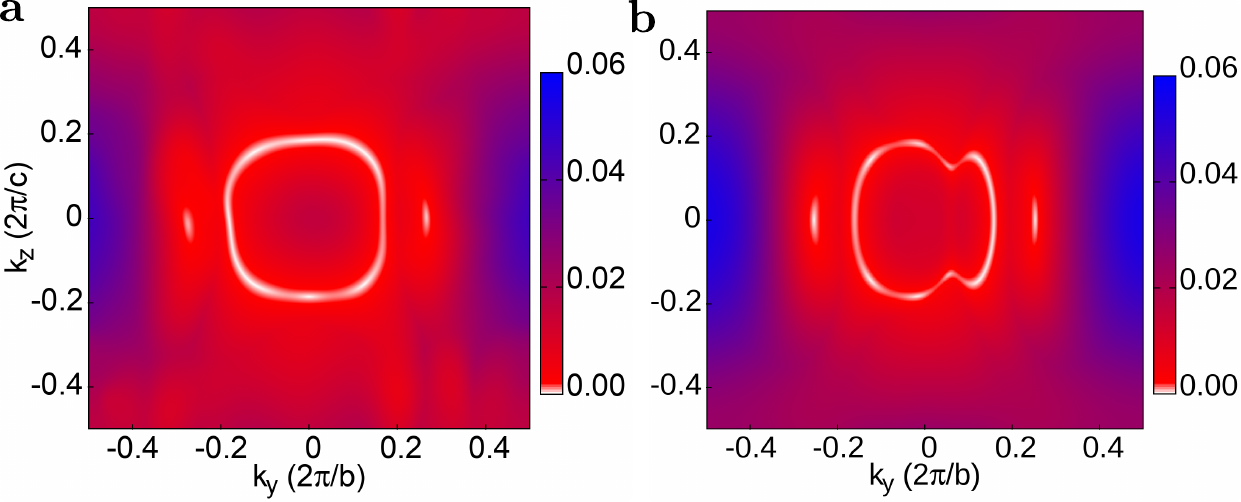}
\caption{The nodal ring centered at the $\Gamma$ point in the $k_x = 0$ plane for a) TiIrGe and b) HfIrGe, calculated without SOC.}
\label{fig:nodal_loop}
\end{figure}

To identify the nodal loop, we employed the WANNIER TOOLS package to compute the local energy gap between the two crossing bands. Our result reveals a symmetry-protected nodal ring encircling the $\Gamma$ point within the $k_x = 0$ plane when SOC is neglected. Figures~\ref{fig:nodal_loop}a,b illustrate the nodal loops of TiIrGe and HfIrGe, respectively. This topological feature is preserved by the glide mirror symmetry $\mathsf{G}_x:(x,y,z)\rightarrow(-x+\frac{1}{2},y+\frac{1}{2},z+\frac{1}{2})$.

\begin{figure*}[t]
\includegraphics[width=0.6\columnwidth]{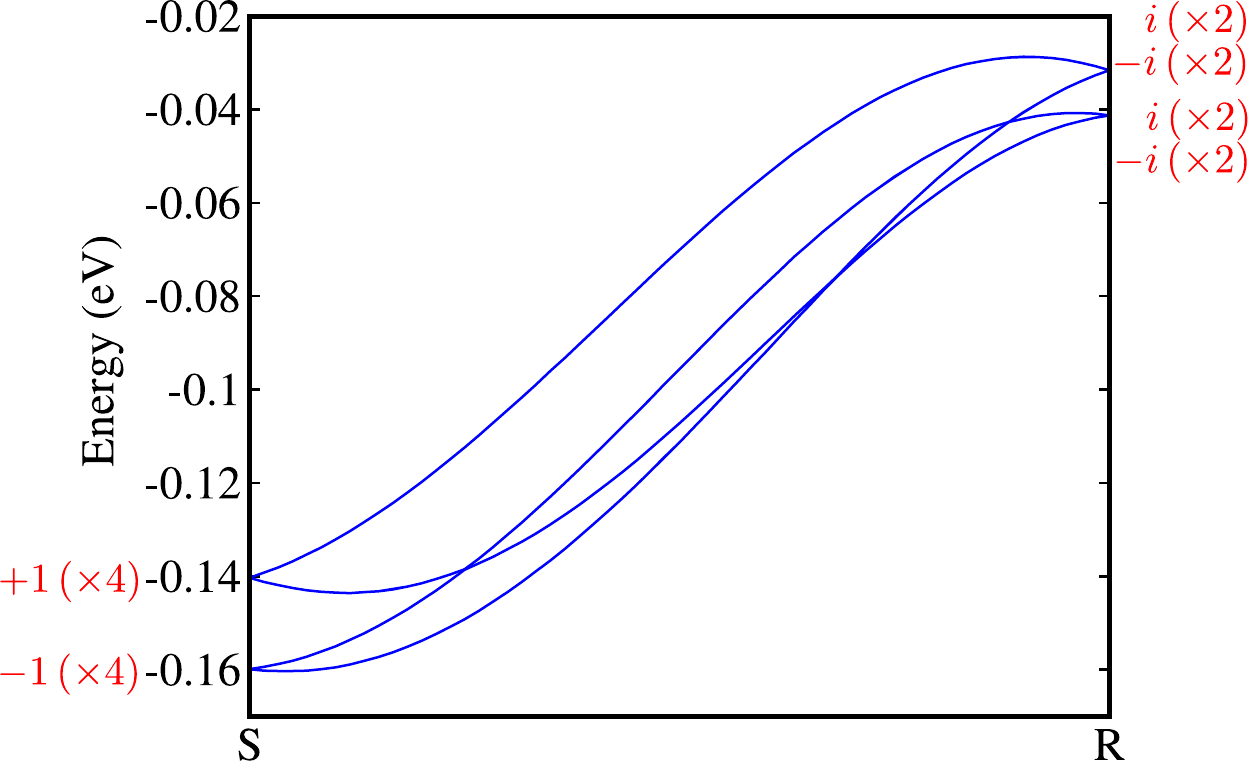}
\caption{Hourglass-type band dispersion of TiIrGe along the S-R path in the presence of spin-orbit coupling (SOC). The numbers and signs in the figure represent the eigenvalues of the glide mirror symmetry operator, $\mathsf{G}_x$.}
\label{fig:Hourglass}
\end{figure*}

Figure~\ref{fig:Hourglass} shows the hourglass-type dispersion of TiIrGe along the S–R path, with eigenvalues of $\mathsf{G}_x$ indicated. At the S point $(\pi,\pi,0)$, each Bloch state $\ket{\phi_n}$ forms a degenerate quartet together with its time-reversal $\mathcal{T} \ket{\phi_n}$, parity $\mathcal{P} \ket{\phi_n}$, and Kramer's $\mathcal{PT} \ket{\phi_n}$ partners. The glide mirror eigenvalues at S are $g_x=\pm 1$, with the $g_x=-1$ states lying at a lower energy than those with $g_x=+1$. At the R point $(\pi, \pi, \pi)$, the glide mirror eigenvalues change to $g_x=\pm i$, meaning that each Kramers pair has opposite eigenvalues. This results in a quartet, with two states having $g_x=+i$ and the other two $g_x=-i$. As the bands evolve from S to R, the redistribution of eigenvalues is enforced by symmetry, inevitably leading to a band crossing. This symmetry-protected crossing gives rise to the characteristic hourglass dispersion shown in Figure~\ref{fig:Hourglass}.

\begin{figure*}[b]
\includegraphics[width=0.7\columnwidth]{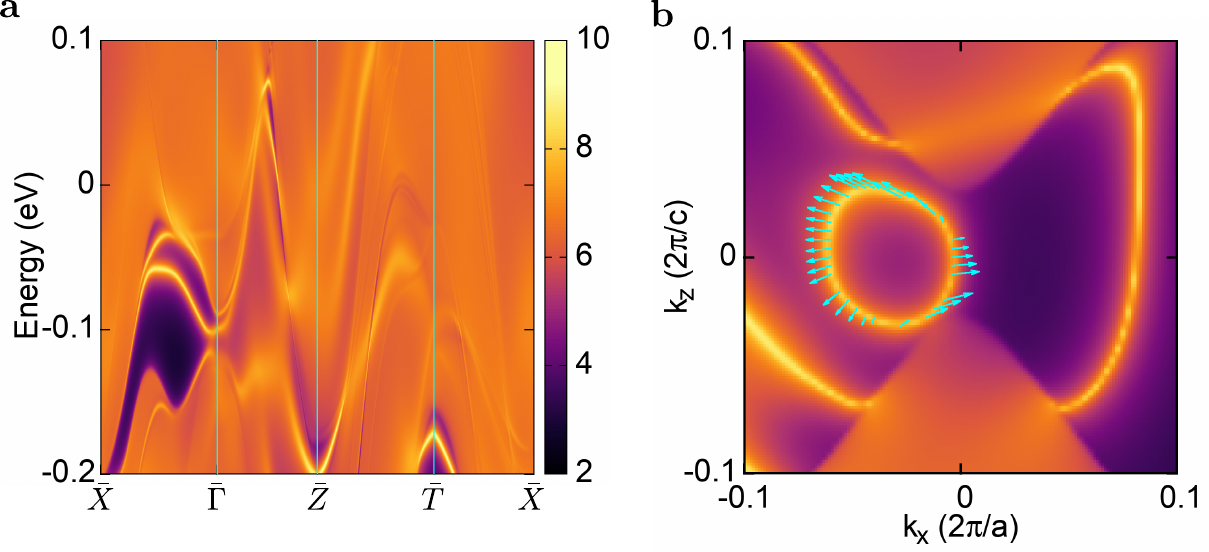}
\caption{a) Surface state spectrum along high-symmetry paths in the projected (010) 2D surface Brillouin zone of HfIrGe. b) Constant energy contour of the spectrum at -0.090 eV for HfIrGe, with cyan arrows indicating the directions of spins.}
\label{fig:Surface_states_z2}
\end{figure*}

The $M$IrGe compounds ($M$= Ti, Hf) lack a global band gap across the Brillouin zone with spin-orbit coupling (SOC). However, the topological invariant $\mathbb{Z}_2$ remains well-defined on the $k_y=0$ plane. Our calculations reveal that $M$IrGe compounds have a topological index of $\mathbb{Z}_2=1$, indicating nontrivial surface states on the (010) surface. Figure~\ref{fig:Surface_states_z2}a clearly shows the topological surface states crossing the Fermi level for HfIrGe for example. Moreover, these states exhibit a helical spin texture as shown in Figure~\ref{fig:Surface_states_z2}b.

\section*{Characterization of $M$I\lowercase{r}G\lowercase{e} samples}

\begin{figure*}[t]
\includegraphics[width=1.0\columnwidth]{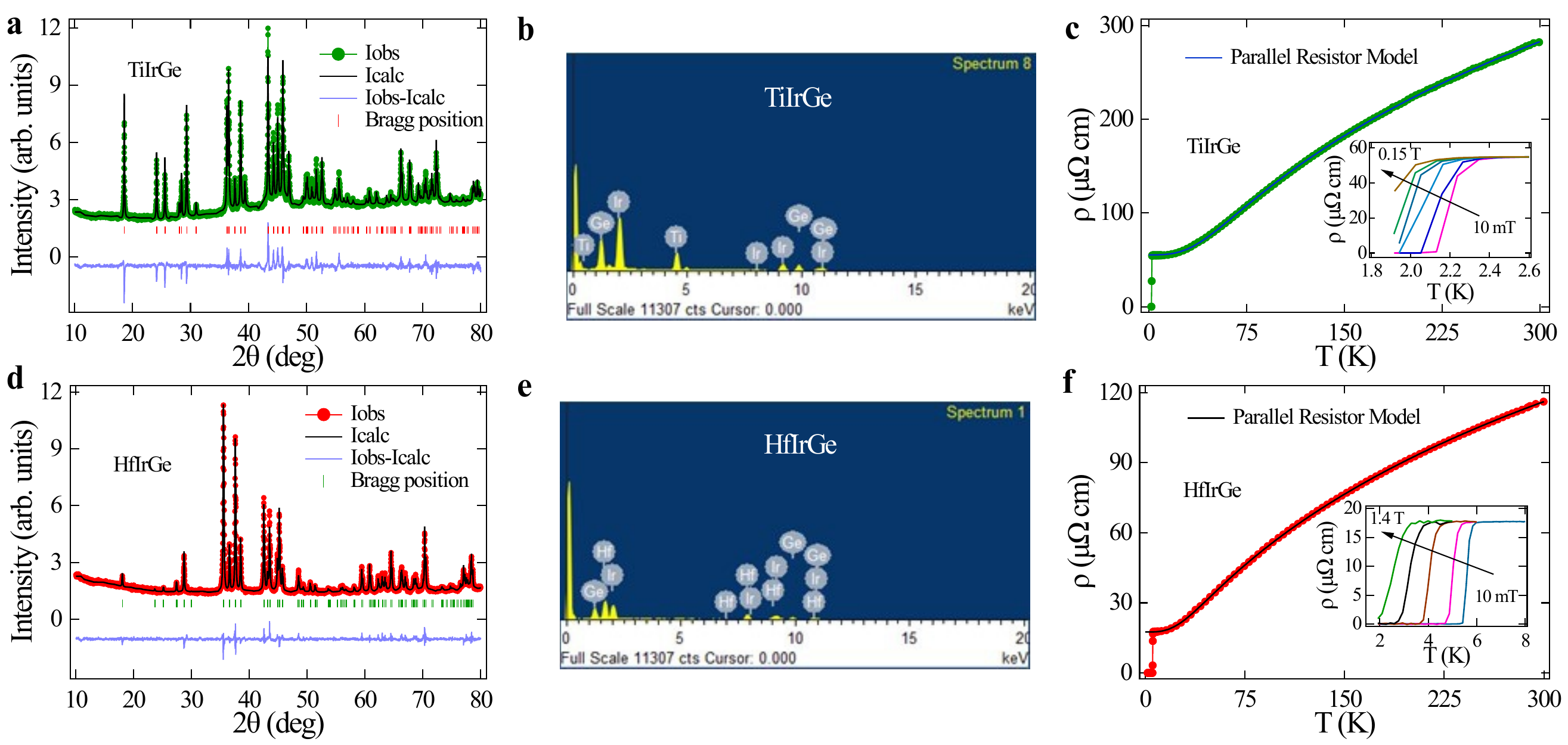}
\caption{a,d) Rietveld-refined powder XRD patterns for the TiIrGe and HfIrGe compounds, in which round markers represent experimental data, while the black lines correspond to the calculated fit. Vertical bars and lines at the bottom denote Bragg's reflection positions and the difference between observed and calculated intensities for both compounds. b,e) Energy-dispersive X-ray spectroscopy (EDAX) spectra for TiIrGe and HfIrGe compounds, respectively. c,f) Temperature-dependent electrical resistivity measured in zero magnetic field, fitted using the parallel resistor model in the normal state, with insets showing resistivity behavior under various applied magnetic fields.}
\label{fig:XRD}
\end{figure*}

\textbf{Structural characterization:} Room temperature powder X-ray diffraction (XRD) measurements of polycrystalline $M$IrGe ($M$ = Ti, Hf) samples were carried out using a PANalytical X'pert Pro diffractometer machine equipped with Cu$K_{\alpha}$ radiation ($\lambda$ = 1.5406 $\text{\AA}$). Rietveld refinement of the powder XRD patterns, conducted using FullProf Suite software \cite{rodriguez1993recent}, confirm that the samples crystallized in the orthorhombic TiNiSi crystal structure. The refined XRD patterns are shown in Figure~\ref{fig:XRD}a,d for TiIrGe and HfIrGe. The obtained lattice parameters are $a$ = 6.2668(9), $b$ = 3.9444(9), and $c$ = 7.3618(4) $\text{\AA}$ and V$_{cell}$ = 181.98(3) $\text{\AA}^{3}$ for TiIrGe, and $a$ = 6.4902(3), $b$ = 4.0136(8), and $c$ = 7.5196(7) $\text{\AA}$ and V$_{cell}$ = 195.88(5) $\text{\AA}^{3}$ for HfIrGe. These values are consistent with the previous report \cite{wang1987_crystal}, as detailed with Wyckoff position in Table \ref{tbl: lattice parameters}. The elemental composition determined by the EDS spectrum as shown in Figure~\ref{fig:XRD}b,e for both polycrystalline compounds suggests a nominal composition.\\

\begin{table}
\caption{Structural parameters of $M$IrGe ($M$ = Ti and Hf) obtained from Rietveld Refinement of powder XRD patterns.}
\label{tbl: lattice parameters}
\setlength{\tabcolsep}{20pt}
\begin{center}
\begin{tabular}[b]{l c c c}
\hline 
\hline
Parameters & TiIrGe & HfIrGe \\
\hline
$a$ ($\text{\AA}$) & 6.2668(9)  & 6.4902(3) \\
$b$ ($\text{\AA}$) & 3.9444(9) & 4.0136(8) \\
$c$ ($\text{\AA}$) & 7.3618(4)  & 7.5196(7) \\
V$_{cell}$ ($\text{\AA}^{3}$)&  181.98(3) &195.88(5)\\
\hline
\end{tabular}
\par\medskip\footnotesize
\end{center}
\begin{center}
\begin{tabular}[b]{l c c c c}\hline 
Atom & Wyckoff position & $x$ & $y$ & $z$\\
\hline
Ti/Hf & 4c & 0.027 & 0.25 & 0.675\\
Ir & 4c & 0.148 & 0.25 & 0.064 \\
Ge & 4c & 0.258 & 0.25 & 0.376\\
\hline
\end{tabular}
\par\medskip\footnotesize
\end{center}
\end{table}

\textbf{Electrical resistivity:} The temperature-dependent resistivity $\rho(T)$ shown in Figure~\ref{fig:XRD}c,f for TiIrGe and HfIrGe compounds exhibit metallic behavior. The obtained residual resistivity ratios (RRR = $\rho_{300K}/\rho_{10K}$) of 5.14(6) and 6.52(7) for TiIrGe and HfIrGe, respectively, indicate high sample quality. A sudden drop in $\rho(T)$ at a transition temperature $T_c$ of 2.24(5) K for TiIrGe and 5.64(4) K for HfIrGe confirms superconductivity. In addition to the zero-field measurements, resistivity measurements at different applied magnetic fields, as shown in the inset of Figure~\ref{fig:XRD}c,f for TiIrGe and HfIrGe were also performed to evaluate the upper critical field, discussed in the next section. Normal state $\rho(T)$ behavior is well-fitted with the theoretical Wiesmann parallel resistivity model given as \cite{wiesmann1977simple};
\begin{equation}
\frac{1}{\rho(T)} = \frac{1}{\rho_{s}} + \frac{1}{\rho_{i}(T)},
\label{Eq1:Parallel}
\end{equation}
where $\rho_{s}$ denotes the temperature-independent saturated resistivity corresponding to a mean free path of the order of the interatomic spacing \cite{fisk1976saturation}. According to Matthiessen's rule, the ideal contribution to resistivity $\rho_{i}(T)$ can be written as the sum of two terms: $\rho_{i}$ = $\rho_{i,0}$ + $\rho_{i, L}$. Here, $\rho_{i,0}$ represents the residual resistivity measured due to impurity scattering, and $\rho_{i, L}$ represents a temperature-dependent contribution of resistivity caused by thermally excited phonons, which can be written as,
\begin{equation}
{\rho_{i,L}(T)}=C \left(\frac{T}{\Theta_{D}}\right)^p \int_{0}^{{\Theta_{D}}/T} \frac{x^{p}}{(e^{x}-1)(1-e^{-x})} dx,
\label{Eq2:Parallel}
\end{equation}
here, $\Theta_{D}$ signifies the Debye temperature, and $C$ is a material-dependent constant. The exponent $p$ can vary depending on the interaction type and is typically set to 3 according to Wilson's theory and 5 as per the BG formula discussed in ref.~\cite{grimvall1981electron}. The best fit of the data for $p$ = 3 (Wilson's theory) yielded a Debye temperature $\Theta_{R}$ = 261(5) K, $\rho_{0}$ = 59.7(7) $\mu\Omega$-cm and $\rho_{sat}$ = 733.4(2) $\mu\Omega$-cm for TiIrGe and $\Theta_{R}$ = 184(9) K, $\rho_{0}$ = 18.7(2) $\mu\Omega$-cm and $\rho_{sat}$ = 272.4(6) $\mu\Omega$-cm for HfIrGe compound, respectively, which are comparable to the structurally similar MIrSi compounds \cite{kase2016superconductivity}.\\

\begin{figure*}[t] 
\includegraphics[width=1.0\columnwidth]{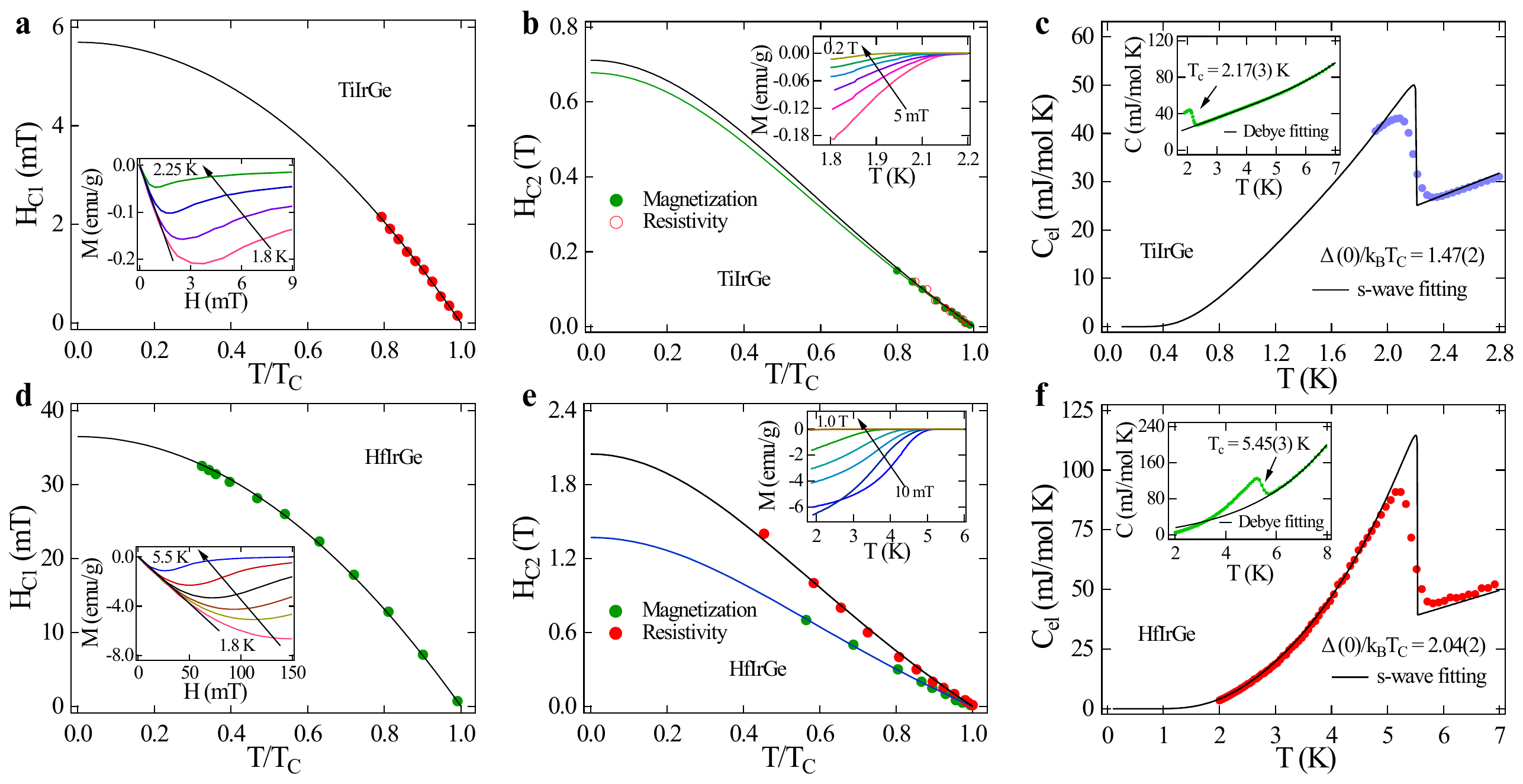}
\caption{a,d) Temperature dependence of the lower critical field fits with the GL relation, with the inset showing a magnetization curve at different temperatures. b,e) The upper critical field was estimated by fitting the resistivity and magnetization data through the GL relation, with the inset showing magnetization curves at different applied fields. c,f) Temperature-dependent total specific heat fitted using the Debye model for TiIrGe and HfIrGe. respectively.}
\label{fig:critical fields_SM}
\end{figure*}

\textbf{Magnetization:} Further superconductivity was confirmed through magnetic susceptibility and found to be diamagnetic at the superconducting transition temperature for both compounds. A weaker diamagnetic signal for field-cooled cooling (FCC) compared to zero-field cooled warming (ZFCW), suggests flux pinning and reveals type-II superconductivity (see main paper). To calculate the lower critical field $H_{c1}(0)$, the magnetic field-dependent magnetization $M(H)$ at various temperatures in the superconducting state was measured, as shown in the inset of Figure~\ref{fig:critical fields_SM}a,d for TiIrGe and HfIrGe compounds. The deviation of the $M(H)$ curve from the Meissner linear line at low fields identifies $H_{c1}$ for each temperature. The Ginzburg-Landau (GL) relation $H_{c1}(T)=H_{c1}(0) [1-\left(T/T_{c}\right)^{2}]$ best fits the extracted temperature dependent $H_{c1}$ data as shown in the Figure~\ref{fig:critical fields_SM}a,d yielding $H_{c1}$(0) values of 5.6(1) and 36.4(1) mT. The high $H_{c1}$ values of HfIrGe compared to the TiIrGe may be influenced by grain boundaries in the polycrystalline samples. In addition, the upper critical field $H_{c2}$(0) is extracted from temperature-dependent magnetization and resistivity measurements under varying magnetic fields. Increasing the magnetic field reduces $T_c$ in the inset of Figure~\ref{fig:critical fields_SM}b,e (similarly in the resistivity curve). The extracted H$_{c2}(T)$ in Figure~\ref{fig:critical fields_SM}b,e were best fitted with the GL relation $H_{c2}(T) = H_{c2}(0)\left[\frac{(1-t^{2})}{(1+t^{2})}\right]$ where $t = \frac{T}{T_{c}}$, providing $H_{c2}$(0) as 0.68(1) and 1.36(1) T from magnetization and 0.71(1) and 2.04(1) T from resistivity measurements for TiIrGe and HfIrGe compounds, respectively.

Superconductivity can be destroyed in two ways under an applied magnetic field: (i) the orbital limiting effect and (ii) the Pauli spin paramagnetic limiting effect. In the orbital limiting effect, the magnetic field increases the kinetic energy of the Cooper pairs, surpassing the condensation energy and leading to the breakdown of superconductivity. For type-II superconductors, the orbital critical field, $H_{c2}^{orb}(0)$, is expressed as $H^{orb}_{c2}(0) = -\alpha T_{c} \left.{\frac{dH_{c2}(T)}{dT}}\right|_{T=T_{c}}$, where $\alpha$ is the purity factor~\cite{werthamer1966temperature, helfand1966temperature}. For $\alpha$ = 0.693, the calculated values of $H^{orb}_{c2}$(0) are 0.53(2) and 1.12(6) T for TiIrGe and HfIrGe compounds, respectively. The Pauli paramagnetic limiting, $H^P_{c2}$(0) field disrupts singlet pairing by aligning electron spins with the applied field. The expression is given as $H^P_{c2}$(0) = 1.86 $T_{c}$ \cite{chandrasekhar1962note, clogston1962upper}, resulting in values of 4.17(5) and 10.49(7) T for the compounds TiIrGe and HfIrGe, respectively ($T_c$ used by magnetization). The relative significance of these two limiting mechanisms in suppressing superconductivity is quantified by the Maki parameter, $\alpha_{m}= \sqrt{2} H_{c2}^{orb}/H_{c2}^{P}$ \cite{maki1966effect}. The calculated $\alpha_{m}$ values are 0.18(1) and 0.15(1) for TiIrGe and HfIrGe compounds, respectively, indicating a minor influence of the Pauli limiting effect in these materials.

Consequently, the coherence length $\xi_{GL}$(0) calculated using the GL formula $H_{c2}(0) = \frac{\Phi_{0}}{2\pi\xi_{GL}^{2}}$ (where $\Phi_{0}$ = $2.07 \times 10^{-15}$ Wb is the flux quantum), yielding values of 22.0(2) and 15.5(6) nm (from magnetization) and 21.5(4) and 12.7(2) nm (from resistivity) for TiIrGe and HfIrGe~\cite{tinkham2004introduction}. The superconducting penetration depth $\lambda_{GL}$(0) is obtained from $H_{c1}$(0) and $\xi_{GL}$(0) using the relation: $H_{c1}(0) = \frac{\Phi_{0}}{4\pi\lambda_{GL}^2(0)}\left[ln \frac{\lambda_{GL}(0)}{\xi_{GL}(0)} + 0.12\right]$, resulting in $\lambda_{GL}$(0) values of 279.1(5) and 92.6(3) nm for TiIrGe and HfIrGe, respectively. Correspondingly, the GL parameters $\kappa_{GL}$ = $\lambda_{GL}(0)$/$\xi_{GL}(0)$ are 12.6(8) and 5.9(5) for TiIrGe and HfIrGe compounds, indicating they are strong type-II superconductors, as $\kappa_{GL}$ exceeds 1/$\sqrt{2}$. Using these parameters and the relation $H_{c}^{2}.ln \kappa_{GL}= H_{c1} H_{c2}$, the thermodynamics critical magnetic field value H$_{c}$ of 38.7(3) and 166.6(2) mT was estimated for TiIrGe and HfIrGe, respectively.\\

\textbf{Specific heat:} 
The temperature-dependent specific heat measured in a zero magnetic field confirms bulk superconductivity in both compounds, with a superconducting jump observed at $T_c$ = 2.17(5) K and 5.45(3) K for TiIrGe and HfIrGe, respectively, as shown in the inset of Figure~\ref{fig:critical fields_SM}c,f. The $T_{c}$ values are consistent with values from other measurements. The normal state data was fitted with the Debye-Sommerfeld relation $C = \gamma_{n} T + \beta_{3} T^{3}$, where $\gamma_{n} T$ is an electronic contribution and $\beta_{3} T^{3}$ indicates the lattice contributions. The best fit to the experimental data provides the Sommerfeld coefficient $\gamma_{n}$ = 9.96(5) mJ/mol-K$^{2}$, Debye constant $\beta_{3}$ = 0.09(1) mJ/mol-K$^{4}$ for TiIrGe, and $\gamma_{n}$ = 7.44(3) mJ/mol-K$^{2}$ and $\beta_{3}$ = 0.21(3) mJ/mol-K$^{4}$ for HfIrGe. The Debye temperature, $\theta_{D}$, was calculated using the relation, $\theta_{D} = \left(\frac{12\pi^{4} R N}{5 \beta_{3}}\right)^{1/3}$, where R = 8.314 J mol$^{-1}$ K$^{-1}$ is the gas constant and N = 3 is the number of atoms per formula unit in $M$IrGe ($M$ = Ti, Hf). From this relation, $\theta_{D}$ was determined to be 399(5) K for TiIrGe and 301(2) K for HfIrGe. Furthermore, the density of state at the Fermi level $D_{C}(E_{F})$ for TiIrGe and HfIrGe compounds are calculated as 4.21(6) and 3.15(6) states $eV^{-1}$ $f.u.^{-1}$ using the relation $\gamma_{n}$ = $\left(\frac{\pi^{2} k_{B}^{2}}{3}\right) D_{C}(E_{F})$, where $k_{B}$ = 1.38 $\times$ 10$^{-23}$ J K$^{-1}$. The electron-phonon coupling, $\lambda_{e-ph}$, is then calculated from McMillan's equation \cite{mcmillan1968transition};
\begin{equation}
\lambda_{e-ph} = \frac{1.04+\mu^{*}\mathrm{ln}(\theta_{D}/1.45T_{c})}{(1-0.62\mu^{*})\mathrm{ln}(\theta_{D}/1.45T_{c})-1.04 };
\label{eqn8:Lambda}
\end{equation}
here, $\mu^{*}$ represents the repulsive screened Coulomb interaction (ranging from 0.07 to 0.15) and is set to 0.13 (for intermetallic compounds). Based on the estimated $\theta_{D}$ and observed $T_{c}$ values, the calculated $\lambda_{e-ph}$ values are 0.48(9) and 0.65(7) for TiIrGe and HfIrGe compounds, suggesting weak coupling superconductivity.

Temperature-dependent electronic specific heat $C_{el}(T)$ is obtained by subtracting the phononic part from the total specific heat, which can be related to the entropy $S$ via the relation: $C_{el} = t\frac{dS}{dt}$. The entropy $S$ for a single-gap BCS superconductor can be expressed as \cite{padamsee1973quasiparticle},
\begin{equation}
\frac{S}{\gamma_{n} T_{c}}= -\frac{6}{\pi^{2}} \left(\frac{\Delta(0)}{k_{B} T_{C}}\right) \int_{0}^{\infty} \left[ {fln(f)+(1-f)ln(1-f)} \right] dy.
\label{BCS}
\end{equation}
Here, $f(\xi)$ = (${e^{E(\xi)/k_{B}T}+1})^{-1}$ is the Fermi function, whereas the integration variable is $y$ = $\xi/\Delta(0)$. $E(\xi)$ is expressed as $\sqrt{\xi^{2} + \Delta^{2}(t)}$ which represents the energy of normal electrons relative to Fermi energy, and $\Delta(t)$ is the temperature-dependent gap function, where t = $T/T_{c}$ is the reduced temperature. The gap function in the isotropic s-wave BCS approximation is expressed as $\Delta(t) = \tanh[1.82(1.018((1/t)-1))^{0.51}]$. The low-temperature $C_{el}(T)$ plot aligns well with the isotropic s-wave model in weak-coupling superconductivity, as shown in the Figure~\ref{fig:critical fields_SM}c,f providing superconducting gap values $\Delta(0)/k_{B}T_{c}$ = 1.47(2) and 2.04(2) for TiIrGe and HfIrGe compounds. A similar family of compounds, such as MIrSi, also exhibits weak coupling superconductivity. However, the fitting for the TiIrGe is not determined accurately due to insufficient data points at low temperatures, which is attributed to the instrument's limitation, extending only up to 1.9 K.

\section*{Analysis of the $\mu$SR results}
\begin{figure*}[!htb]
\includegraphics[width=0.8\columnwidth]{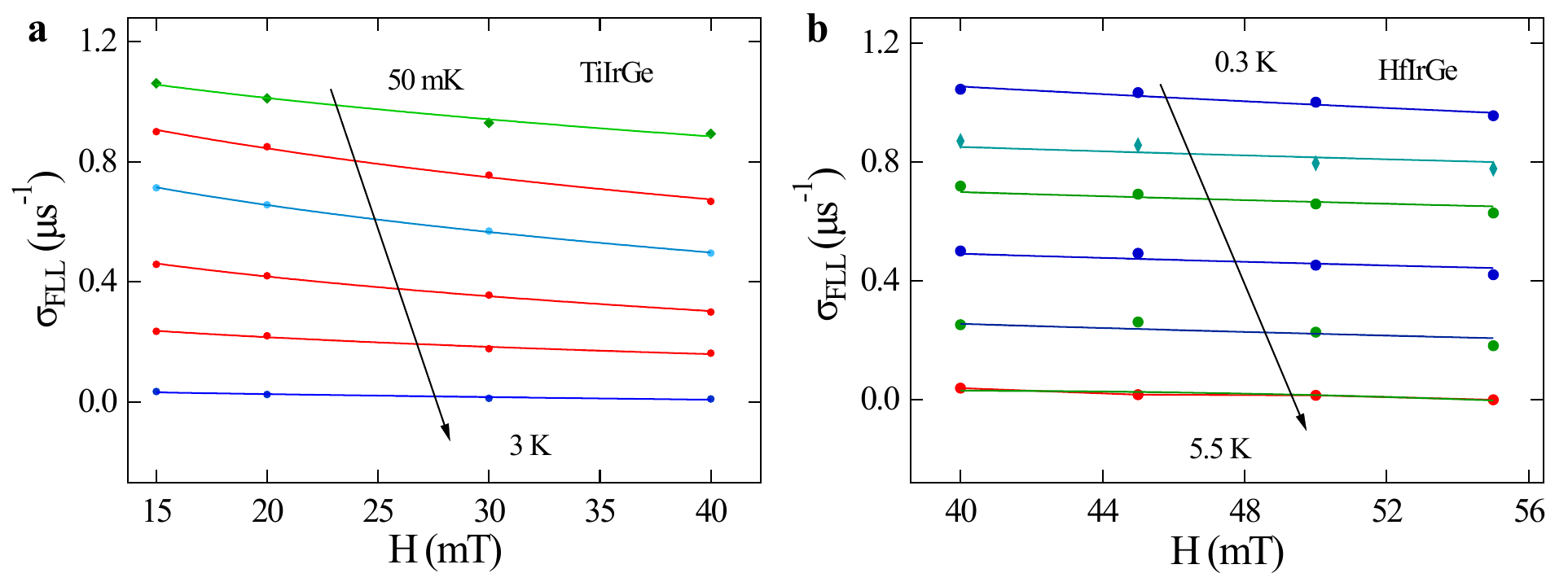}
\caption{a,b) are the field-dependent relaxation at various temperatures for TiIrGe and HfIrGe compounds, respectively.}
\label{fig:sigmawithfield_SM}
\end{figure*}

\begin{figure*}[!htb]
\includegraphics[width=0.8\columnwidth]{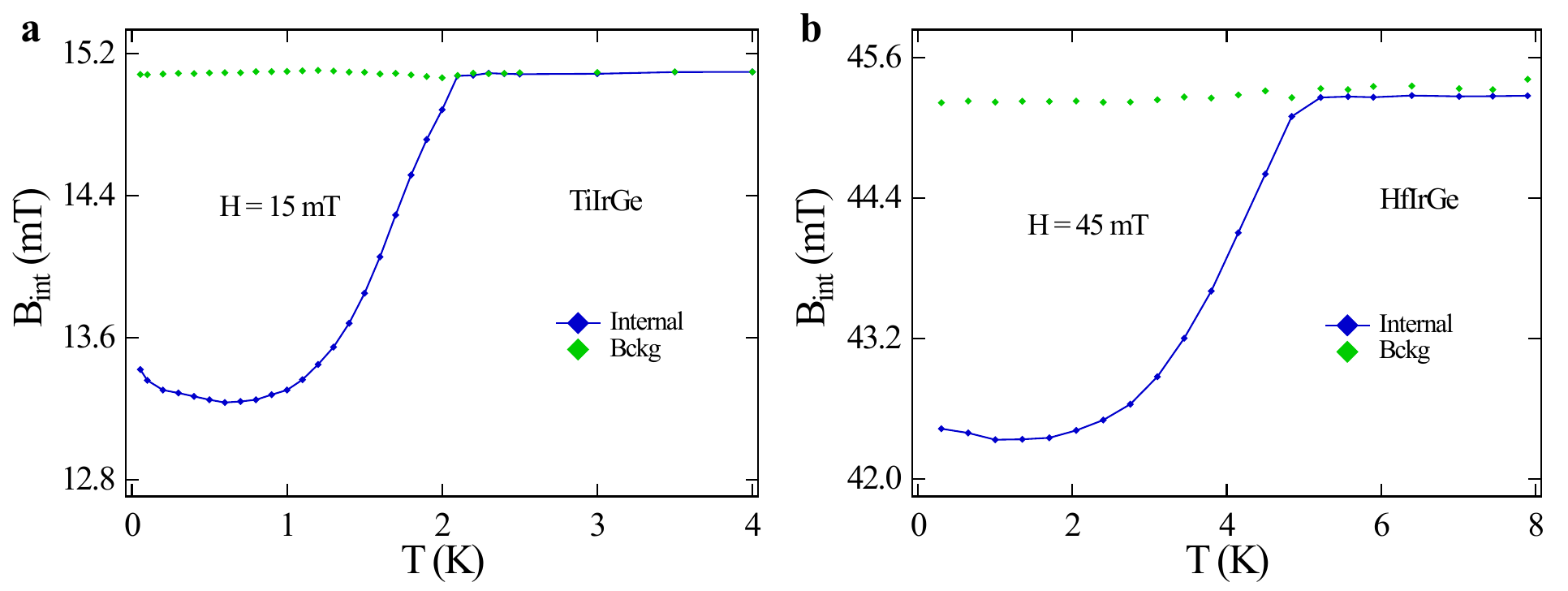}
\caption{a,b) are the temperature-dependent internal magnetic fields for TiIrGe and HfIrGe compounds, respectively.}
\label{fig:internaldield_SM}
\end{figure*}

The field-dependent relaxation rate at various temperatures for TiIrGe and HfIrGe compounds are shown in Figure~\ref{fig:sigmawithfield_SM}a,b. These measurements were used to calculate the penetration depth at low values of the upper critical field for the compounds, with the relation and calculated results provided in the main paper. Figures~\ref{fig:internaldield_SM}a,b show the temperature-dependent internal magnetic field at an applied magnetic field of 15 and 45 mT, for TiIrGe and HfIrGe, respectively. In the superconducting state, the internal magnetic field is less than the applied field due to Meissner field expulsion, but above $T_c$, it matches the applied field and overlaps with the background magnetic contribution, which is constant over the measured temperature range.

\begin{figure}
\includegraphics[width=0.5\columnwidth]{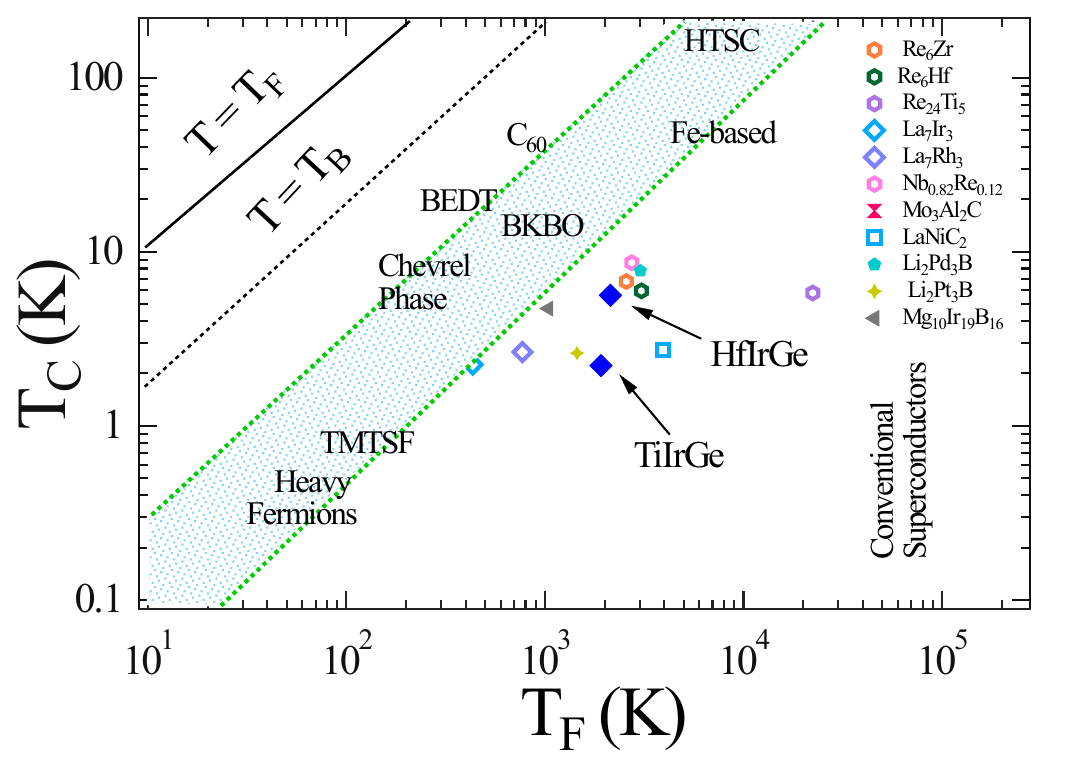}
\caption {\label{Fig6} The Uemura plot classifies the superconductor in conventional and unconventional types based on $T_{C}$ and $T_{F}$. $M$IrGe ($M$ = Ti, Hf) is positioned as a blue marker near the conventional range of superconductivity.}
\end{figure}

\begin{table}
\caption{Superconducting and normal parameters of $M$IrGe ($M$ = Ti and Hf) determined from various measurements, including magnetization, resistivity, specific heat, and $\mu$SR.}
\label{tbl: parameters}
\setlength{\tabcolsep}{0pt}
\begin{center}
\begin{tabular}{p{0.15\linewidth}p{0.15\linewidth}p{0.15\linewidth}p{0.15\linewidth}p{0.15\linewidth}}
\hline
\hline
Parameters & Units & TiIrGe & HfIrGe & IrGe \cite{arushi2022microscopic}\\
\hline
T$_{c}$ & K & 2.24(5) & 5.64(4) &4.74\\
H$_{c1}(0)$ & mT & 5.6(1)  &36.4(1) &13.3(2)\\ 
H$_{c2}^{mag}$(0) & T & 0.68(1)  &1.36(1) &1.13(2)\\
H$_{c2}^{res}$(0) & T & 0.71(1)  &2.04(1) &-\\
H$_{c}$ & mT & 38.7(3) &  166.6(2) &78\\
$\xi_{GL}^{mag}$& nm & 22.0(2)  &15.5(6) &17.0 \\
$\lambda_{GL}^{mag}$& nm &279.1(5) &92.6(3) &175\\
$\lambda_{GL}^{\mu SR}$& nm & 273.1(6) & 246.1(9) &134\\
$k_{GL}$& & 12.6(8)  &5.9(5) &10\\
$\gamma_{n}$& mJ/(mol K$^{2}$) & 9.96(5) & 7.44(4) & 3.1\\
$\theta_{D}$& K& 399(5) & 301(2) &160(1) \\
$\frac{\Delta(0)}{k_{B}T_{C}}$ (sp) &   & 1.47(2) &2.04(2) & 2.3(1)\\
${\Delta^{\mu SR}(0)}$ & meV  & 0.30(9) & 0.75(2) &-\\
$\frac{\Delta^{\mu SR}(0)}{k_{B}T_{C}}$ & & 1.66(7) & 1.68(2) &2.1\\
$\lambda_{e-ph}$ &  & 0.48(9) & 0.65(7)&0.78(2)\\
D$_{C}$(E$_{F}$) & states/(eV f.u.) & 4.21(6) & 3.15(6)&- \\
$n$& $10^{26}$ &5.6(2)&7.7(3)&-\\
$m^{*}$ & $m_{e}$ &1.48(9)&1.65(7)&-\\
$T_{F}$& K &1921(14)&2136(19)&-\\
\hline
\hline
\end{tabular}
\par\medskip\footnotesize
\end{center}
\end{table}

Here, we estimate the superconducting carrier density $n_s$  = 5.62(4) $\times 10^{26}$ and 7.7(3) $\times 10^{26}$ $m^{-3}$ by $n_s(0) = m^*/ \mu_{0} e^2 \lambda^2$, where $m^* = (1+\lambda_{e-ph}) m_{e}$. We estimated the Fermi temperature ($T_{F}$) for TiIrGe and HfIrGe compounds using the given equation \cite{hillier1997classification}, where $n$ and $m^{*}$ are the electronic carrier density and the effective mass of quasi-particles, respectively:
\begin{equation}
 k_{B}T_{F} = \frac{\hbar^{2}}{2}(3\pi^{2})^{2/3}\frac{n^{2/3}}{m^{*}}, 
\label{eqn12:tf}
\end{equation}

The obtained values of $T_{F}$ are 1921(14) and 2136(19) K, using $\lambda$ =  2731(6) and 2461(9) $\text{\AA}$ (from muon spectroscopy measurements) and $\lambda_{e-ph}$ = 0.48(9) and 0.65(7) (from specific heat data) for TiIrGe and HfIrGe, respectively. The classification of Uemura can divide superconductors as conventional and unconventional based on the ratio of transition temperature and Fermi temperature (T$_{C}$/T$_{F}$) \cite{uemura1989universal, uemura1991basic}. If this value comes in the range 0.01 $\le$ $\frac{T_{C}}{T_{F}}$ $\le$ 0.1, then a superconductor is considered an unconventional superconductor; high-T$_{C}$ superconductors, heavy fermion superconductors, Fe-based superconductors, as well as Chevrel phase superconductors lie inside this unconventional band on the Uemura plot. The ratios of $T_{C}$ to $T_{F}$ are 0.0011(5) and 0.0026(3) by using T$_{c}$ value of 2.22 and 5.62 K for TiIrGe and HfIrGe compounds, marked by a blue marker in Figure~\ref{Fig6} for $M$IrGe ($M$ = Ti, Hf), indicating its location significantly outside the region of unconventional superconductors.

\bibliographySupp{Houglass_suppl_refs}

\end{document}